\documentclass[aps,reprint,superscriptaddress,pra]{revtex4-2}

\usepackage{amsmath,amssymb,amsthm}
\usepackage{booktabs}
\usepackage{float}
\usepackage{graphicx}
\usepackage{microtype}
\usepackage{parskip}
\usepackage[section]{placeins}
\usepackage[
    colorlinks=true,
    linkcolor=blue,
    citecolor=blue,
    urlcolor=blue
]{hyperref}
\usepackage{bm}
\usepackage{listings}
\usepackage{xcolor}
\usepackage{braket}

\usepackage{tikz}
\usetikzlibrary{positioning}

\begin{document}

\title{\textbf{Worldline-Susceptibility Scheduling for Quantum Annealing Beyond Local-Adiabatic Evolution}}


\author{Suraj Singh}
\thanks{These two authors contributed equally to this work.}
\affiliation{The Institute of Mathematical Sciences (IMSc), C.I.T Campus, Taramani, Chennai 600113, India}
\affiliation{QCAR Group, The Institute of Mathematical Sciences (IMSc), Chennai 600113, India}
\affiliation{Pecslab Research}

\author{Lakshya Nagpal}
\thanks{These two authors contributed equally to this work.}
\affiliation{The Institute of Mathematical Sciences (IMSc), C.I.T Campus, Taramani, Chennai 600113, India}
\affiliation{QCAR Group, The Institute of Mathematical Sciences (IMSc), Chennai 600113, India}
\affiliation{Pecslab Research}

\author{Vikas Chauhan}
\affiliation{Department of Physics, Ramjas College, University of Delhi, Delhi 110007, India}
\affiliation{QCAR Group, The Institute of Mathematical Sciences (IMSc), Chennai 600113, India}

\author{S.~R.~Hassan}
\affiliation{The Institute of Mathematical Sciences (IMSc), C.I.T Campus, Taramani, Chennai 600113, India}
\affiliation{QCAR Group, The Institute of Mathematical Sciences (IMSc), Chennai 600113, India}
\affiliation{Homi Bhabha National Institute, Anushakti Nagar, Mumbai, Maharashtra 400094}

\begin{abstract}
The performance of quantum annealing depends critically on how the available annealing time is distributed along the evolution. Although the Roland--Cerf local-adiabatic schedule is theoretically optimal, it requires complete knowledge of the instantaneous spectral gap, making it impractical for large optimization problems. We propose a computationally inexpensive surrogate schedule based on the worldline magnetization susceptibility measured during simulated quantum annealing. The susceptibility is obtained directly from equilibrium Monte Carlo sampling and identifies the critical region of the anneal without requiring spectral information. Using exact diagonalization of Sherrington--Kirkpatrick spin-glass instances as ground truth, we show that the resulting schedule consistently outperforms conventional linear annealing and, for a substantial fraction of instances, also surpasses the exact Roland--Cerf schedule. We demonstrate that this unexpected behaviour originates from two finite-time failure modes of exact local-adiabatic scheduling: a boundary-gap trap, in which the minimum spectral gap occurs at the end of the anneal, and an oscillatory instability caused by excessively localized time allocation around an interior minimum gap. These results suggest that robust scheduling based on inexpensive equilibrium observables can outperform exact spectral-gap-based strategies under realistic finite-time conditions. The complete methodology is implemented in the open-source \texttt{Qanneal} framework.
\end{abstract}

\maketitle

\section{Introduction}

Quantum annealing has emerged as a powerful framework for solving combinatorial optimization problems by encoding their solutions into the ground state of an Ising Hamiltonian. A broad range of applications, including scheduling, routing, portfolio optimization, machine learning, and spin-glass optimization, can be formulated within this framework and solved by continuously evolving the system from the ground state of a simple driver Hamiltonian to that of the target problem Hamiltonian \cite{kadowaki1998,farhi2001,santoro2002,lucas2014,albash2018,Cro20b,Kin22,Hau19,Lau14}. The development of dedicated quantum annealing hardware has further stimulated interest in understanding how the annealing dynamics can be optimized to improve the performance of quantum optimization algorithms.

The success of quantum annealing depends not only on the choice of the problem Hamiltonian but also on how the annealing parameter is varied in time. For a fixed total annealing time, traversing the most difficult part of the evolution too rapidly increases the probability of diabatic transitions, whereas evolving slowly throughout the entire annealing path wastes valuable computational resources. The annealing schedule therefore determines how the available runtime is distributed along the evolution and plays a central role in the final ground-state success probability. Designing schedules that allocate computational effort where it is most beneficial provides one of the few opportunities to improve annealing performance without modifying either the optimization problem or the underlying hardware.

Theoretically, this problem has an elegant solution. Roland and Cerf showed that the annealing velocity should satisfy a local adiabatic condition in which the evolution slows down in inverse proportion to the square of the instantaneous spectral gap \cite{roland2002}. Under the assumptions of adiabatic evolution, this local-adiabatic schedule minimizes the total annealing time while maintaining the desired adiabatic accuracy and has therefore become the standard theoretical benchmark for schedule design. Rigorous error bounds justifying the adiabatic approximation on which this condition rests have been established for general gapped Hamiltonians \cite{jansen2007,Lid08}. In practice, however, implementing this prescription requires complete knowledge of the instantaneous spectral gap throughout the annealing trajectory. Since obtaining the low-energy spectrum generally requires repeated diagonalization of an exponentially large Hamiltonian, constructing the exact local-adiabatic schedule rapidly becomes computationally infeasible for optimization problems of practical interest.

This limitation naturally motivates an alternative approach. Rather than attempting to compute the spectral gap explicitly, can one identify an inexpensive observable that reliably locates the critical region of the anneal and therefore provides the essential information required for schedule construction? Simulated quantum annealing (SQA) provides a natural setting for addressing this question. Through the Suzuki--Trotter mapping, the quantum annealing process is represented by an equivalent classical worldline system, allowing equilibrium observables to be obtained efficiently by classical Monte Carlo sampling without requiring access to the quantum spectrum.

In this work, we propose the worldline magnetization susceptibility measured during SQA as a practical surrogate for local-adiabatic scheduling. The susceptibility is obtained directly from equilibrium fluctuations of the worldline configuration and provides a computationally inexpensive indicator of the quantum critical region encountered during the annealing process. Using exact diagonalization of Sherrington--Kirkpatrick spin-glass instances as ground truth, we demonstrate that the resulting surrogate schedules consistently outperform conventional linear annealing and, for a substantial fraction of instances, also outperform the exact Roland--Cerf schedule.

More importantly, our study reveals that this seemingly counterintuitive result is not accidental. We identify two distinct finite-time failure modes of exact local-adiabatic scheduling. The first is a \emph{boundary-gap trap}, in which the minimum spectral gap occurs at the end of the annealing path, causing the schedule to allocate most of the runtime after the transverse-field driver has effectively disappeared. The second is an \emph{oscillatory instability} arising from an excessively localized concentration of annealing time around an interior minimum gap, leading to coherent multilevel interference during finite-time evolution. These results suggest that, under realistic finite-time conditions, robustness of the schedule can be more important than reproducing the exact spectral-gap profile.

To ensure that the proposed methodology is reproducible and readily applicable to future studies, the complete computational workflow has been implemented in the open-source \texttt{Qanneal} framework. The software integrates simulated quantum annealing, worldline susceptibility measurements, surrogate schedule construction, exact spectral benchmarking, and quantum-dynamical simulations within a unified platform. The framework is briefly described in Appendix~\ref{app:qanneal}, while detailed software documentation will accompany the public release of the code.

\section{Surrogate Annealing Schedule Design via Simulated Quantum Annealing}

\subsection{Quantum Annealing and Local-Adiabatic Scheduling}

Quantum annealing solves combinatorial optimization problems by encoding the objective function into the ground state of a problem Hamiltonian and continuously evolving from the ground state of a simple driver Hamiltonian. The evolution is governed by
\begin{equation}
\hat H(s)
=
A(s)\hat H_D
+
B(s)\hat H_P,
\qquad
0\le s\le1,
\label{eq:qa_hamiltonian}
\end{equation}
where $\hat H_D$ denotes the driver Hamiltonian, $\hat H_P$ is the problem Hamiltonian, and the annealing functions satisfy
\begin{equation}
A(0)=1,\qquad
B(0)=0,
\qquad
A(1)=0,\qquad
B(1)=1.
\end{equation}

The system is initialized in the easily prepared ground state of $\hat H_D$. During the anneal, the transverse-field driver is gradually suppressed while the problem Hamiltonian is simultaneously turned on, with the objective of preparing the ground state of $\hat H_P$, which represents the optimal solution of the optimization problem.

The efficiency of this evolution is governed by the instantaneous spectral gap
\begin{equation}
\Delta(s)
=
E_1(s)-E_0(s),
\label{eq:gap}
\end{equation}
where $E_0(s)$ and $E_1(s)$ denote the instantaneous ground- and first-excited-state energies of $\hat H(s)$. When the spectral gap becomes small, the probability of diabatic transitions increases, requiring the evolution to slow down in order to remain close to the instantaneous ground state.

Roland and Cerf showed that this requirement leads to the local-adiabatic scheduling condition
\begin{equation}
\left|
\frac{ds}{dt}
\right|
=
\varepsilon
\frac{\Delta(s)^2}
{\left|
\left<
E_1(s)
\left|
\partial_s\hat H
\right|
E_0(s)
\right>
\right|},
\label{eq:RC}
\end{equation}
where $\varepsilon$ specifies the allowable adiabatic error. This schedule allocates additional annealing time to regions where the spectral gap is small and is widely regarded as the optimal schedule within the local adiabatic approximation \cite{roland2002}.

Despite its theoretical appeal, Eq.~(\ref{eq:RC}) has limited practical applicability. Constructing the schedule requires knowledge of the instantaneous low-energy spectrum throughout the annealing trajectory, including both the spectral gap and the transition matrix element appearing in Eq.~(\ref{eq:RC}). For generic optimization problems, these quantities can only be obtained through repeated diagonalization of an exponentially large Hamiltonian, rendering the exact local-adiabatic schedule computationally inaccessible beyond relatively small system sizes.

This limitation motivates the central objective of the present work. Rather than attempting to compute the instantaneous spectrum explicitly, we investigate whether simulated quantum annealing (SQA) can itself provide sufficient information to construct an effective annealing schedule. Specifically, we seek a computationally inexpensive observable that can be measured directly during an SQA simulation, identifies the critical region of the annealing process, and thereby enables the construction of a practical surrogate to the exact local-adiabatic schedule.

\subsection{Simulated Quantum Annealing as a Surrogate Scheduling Framework}

The central idea of this work is to extract scheduling information directly from a simulated quantum annealing (SQA) simulation, thereby avoiding explicit calculations of the instantaneous quantum spectrum. Rather than reproducing the quantum dynamics in real time, SQA exploits the Suzuki--Trotter mapping \cite{suzuki1976,trotter1959} to transform the transverse-field Ising model into an equivalent classical statistical-mechanical system that can be sampled efficiently using classical Monte Carlo techniques \cite{santoro2002,Ban21b,Mbe18}. This mapping provides access to equilibrium observables that reflect the evolution of the underlying quantum system and therefore offers a practical route to constructing surrogate annealing schedules.

Applying the Suzuki--Trotter decomposition to Eq.~(\ref{eq:qa_hamiltonian}) yields the effective classical action
\begin{equation}
S_{\rm eff}
=
\frac{\beta}{M}
\sum_{k=1}^{M}
H_P
\left(
\mathbf{s}^{(k)}
\right)
-
J_\perp
\sum_{k=1}^{M}
\sum_{i=1}^{N}
s_i^{(k)}
s_i^{(k+1)},
\label{eq:Seff}
\end{equation}
where $M$ denotes the number of Trotter slices, $\beta$ is the inverse temperature, and the effective coupling along the imaginary-time direction is
\begin{equation}
J_\perp(\beta,\Gamma)
=
\frac12
\ln
\left[
\frac{1}
{\tanh(\beta\Gamma/M)}
\right].
\label{eq:Jperp}
\end{equation}

The quantum annealing problem is therefore mapped onto a $(d+1)$-dimensional classical Ising model consisting of $M$ coupled replicas of the original system. The additional imaginary-time dimension encodes the quantum fluctuations generated by the transverse field, while equilibrium configurations of the classical system represent quantum worldline configurations.

Within this worldline representation, we monitor the average magnetization
\begin{equation}
m
=
\frac{1}{NM}
\sum_{i=1}^{N}
\sum_{k=1}^{M}
s_i^{(k)},
\label{eq:m}
\end{equation}
whose equilibrium fluctuations define the worldline magnetization susceptibility,
\begin{equation}
\chi_m
=
NM
\left(
\langle m^2\rangle
-
\langle |m| \rangle^2
\right).
\label{eq:chim}
\end{equation}

The susceptibility follows directly from the fluctuation--dissipation theorem and is obtained with essentially no additional computational cost beyond the Monte Carlo simulation itself. As the transverse field decreases during the anneal, the worldline configurations evolve continuously from a quantum-disordered regime to a classically ordered regime. This crossover is accompanied by a pronounced peak in $\chi_m$, reflecting the enhanced fluctuations that occur near the quantum critical region.

Near the critical region, the worldline susceptibility follows the expected critical scaling,
\begin{equation}
\chi_m(s)
\propto
\frac{1}{\Delta(s)^2},
\label{eq:chi_scaling}
\end{equation}
providing a physically motivated surrogate for the inverse-square spectral gap that governs the Roland--Cerf schedule \cite{You07,Alb09,Pel22,Ovi21}. Consequently, the susceptibility identifies the region where the annealing dynamics is most sensitive without requiring any explicit knowledge of the instantaneous eigenvalue spectrum.

All SQA simulations presented in this work are performed using the open-source \texttt{qanneal} framework. The software implements worldline quantum Monte Carlo simulations, measures the susceptibility during the annealing process, and constructs surrogate annealing schedules directly from the measured equilibrium observables. Implementation details are summarized in Appendix~A.

\subsection{Construction of the Surrogate Annealing Schedule}

The objective of the surrogate scheduling framework is not to reproduce the instantaneous spectral gap itself, but to identify the critical region of the annealing process where additional computational effort is most beneficial. Since the worldline susceptibility exhibits a pronounced maximum in the vicinity of the minimum spectral gap, it provides a natural measure of the relative importance of different stages of the annealing trajectory.

We therefore construct the annealing schedule directly from the susceptibility measured during the SQA simulation. The measured susceptibility is first converted into a positive time-allocation weight,
\begin{equation}
w(s)
=
\chi_m(s)+\chi_0,
\label{eq:weight}
\end{equation}
where $\chi_0>0$ is a small regularization constant introduced to avoid singular behaviour in regions where the susceptibility becomes very small.

The cumulative fraction of the total annealing time assigned up to annealing parameter $s$ is then defined as
\begin{equation}
\tau(s)
=
\frac{
\displaystyle\int_0^{s}
w(s')\,ds'
}{
\displaystyle\int_0^{1}
w(s')\,ds'
},
\label{eq:tau}
\end{equation}
which satisfies
\begin{equation}
\tau(0)=0,
\qquad
\tau(1)=1.
\end{equation}

The function $\tau(s)$ therefore represents the normalized cumulative time allocation along the annealing trajectory. Regions where the susceptibility is large accumulate annealing time more rapidly than regions where the susceptibility is small, naturally concentrating computational effort around the critical region.

The surrogate annealing schedule is obtained by inverting the cumulative mapping,
\begin{equation}
s(t)
=
\tau^{-1}
\left(
\frac{t}{T}
\right),
\qquad
0\le t\le T,
\label{eq:schedule}
\end{equation}
where $T$ denotes the total annealing time.

Differentiating Eq.~(\ref{eq:schedule}) yields the instantaneous annealing velocity,
\begin{equation}
\frac{ds}{dt}
=
\frac{1}
{T\,w(s)}
=
\frac{1}
{T\left[\chi_m(s)+\chi_0\right]}.
\label{eq:velocity}
\end{equation}

The resulting schedule automatically slows the evolution in regions where the worldline susceptibility is large and accelerates it where the susceptibility is small, while preserving the prescribed total annealing time. Unlike the Roland--Cerf schedule, every quantity entering the construction is obtained directly from equilibrium measurements performed during the SQA simulation. Consequently, the computational cost of constructing the schedule scales with the classical Monte Carlo simulation rather than with repeated diagonalization of the exponentially large quantum Hamiltonian.

Within the \texttt{qanneal} framework, this entire workflow---worldline simulation, susceptibility measurement, cumulative time allocation, and surrogate schedule generation---is performed automatically, providing a practical implementation of the surrogate scheduling strategy for optimization problems beyond the reach of exact spectral methods.

\subsection{Computational Workflow and Qanneal Implementation}

The surrogate scheduling methodology consists of four computational stages. First, the optimization problem is encoded as a transverse-field Ising Hamiltonian and the annealing path
\begin{equation}
H(s)=A(s)H_D+B(s)H_P
\end{equation}
is specified. For the small benchmark systems considered in this work, the instantaneous spectral gap is computed by exact diagonalization solely for validation against the surrogate method.

Second, the same annealing path is simulated using the Suzuki--Trotter worldline formulation. At each value of the annealing parameter, equilibrium worldline configurations are generated using quantum Monte Carlo sampling, from which the worldline magnetization susceptibility $\chi_m(s)$ is evaluated through its equilibrium fluctuations.

Third, the measured susceptibility is converted into a cumulative time-allocation function using Eq.~(\ref{eq:tau}), and the corresponding surrogate annealing schedule is obtained by numerical inversion of the cumulative mapping. Since the schedule depends only on equilibrium observables measured during the SQA simulation, no information about the instantaneous eigenvalue spectrum is required.

Finally, the quality of the surrogate schedule is assessed by solving the time-dependent Schr\"odinger equation using identical total annealing times for three different scheduling protocols: the conventional linear schedule, the exact Roland--Cerf schedule, and the proposed SQA-derived surrogate schedule. The performance is quantified by the final ground-state probability,
\begin{equation}
P_{\rm GS}
=
\left|
\langle
\psi_0(T)
|
\psi(T)
\rangle
\right|^2,
\end{equation}
where $\psi(T)$ is the final evolved state and $\psi_0(T)$ is the instantaneous ground state of the problem Hamiltonian.

The complete computational pipeline has been implemented in the open-source \texttt{Qanneal} software package. Qanneal provides an integrated framework for simulated quantum annealing, worldline quantum Monte Carlo simulations, susceptibility measurements, surrogate schedule construction, and benchmarking against exact quantum dynamics. Although exact diagonalization is employed in the present work to validate the surrogate schedules on small benchmark systems, the schedule construction itself relies exclusively on the classical SQA simulation. Consequently, the methodology remains applicable to problem sizes well beyond the reach of exact spectral calculations, making Qanneal a practical platform for scalable surrogate schedule design.

\section{Results}

We now evaluate the worldline-susceptibility schedule against two reference protocols: a uniform linear schedule and the local-adiabatic Roland–Cerf schedule constructed from exact spectral information. Unless otherwise stated, all simulated quantum annealing calculations, susceptibility measurements, and surrogate schedules reported below were generated using the open-source \texttt{Qanneal} framework described in Appendix~A. Exact diagonalization and real-time quantum evolution are employed only to provide spectral benchmarks and to assess the performance of the resulting schedules; neither is required to construct the surrogate schedule itself.

We first examine whether the worldline magnetization susceptibility extracted from SQA identifies the same difficult region of the annealing path as the exact spectral gap. We then compare how the linear, Roland–Cerf, and susceptibility-derived schedules distribute the available annealing time before analysing their finite-time ground-state preparation performance.

\subsection{Worldline Susceptibility as an Indicator of the Minimum-Gap Region}

The proposed construction does not require the worldline susceptibility to reproduce the instantaneous spectral gap quantitatively. It requires only that the susceptibility identify the broad region in which the annealing dynamics becomes most sensitive and where additional evolution time may therefore be beneficial.

\begin{figure*}[t]
\centering
\includegraphics[width=0.78\linewidth]{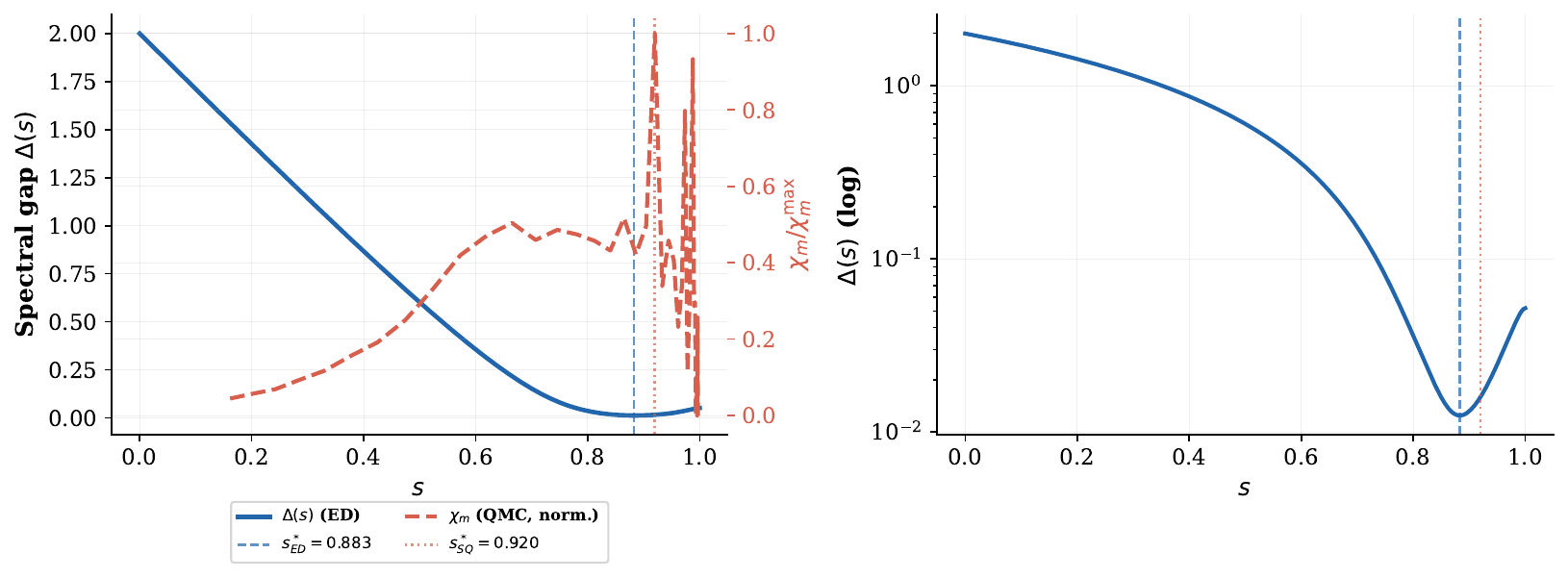}
\caption{
Comparison between the exact instantaneous spectral gap and the normalized worldline magnetization susceptibility measured in SQA for a representative $n=10$ Sherrington–Kirkpatrick instance. The minimum of $\Delta(s)$ occurs at $s_{\rm ED}^{}=0.883$, while the maximum of $\chi_m(s)$ occurs at $s_{\rm SQ}^{}=0.920$. Left: gap and normalized susceptibility over the full annealing path. Right: logarithmic representation of the spectral gap, highlighting the proximity of the two characteristic locations. The susceptibility identifies the minimum-gap region without using spectral information.
}
\label{fig:gap_chiB}
\end{figure*}

We test this premise by comparing the susceptibility measured in SQA with the exact instantaneous spectral gap obtained by diagonalizing the corresponding quantum Hamiltonian. Figure~\ref{fig:gap_chiB} presents this comparison for a representative Sherrington–Kirkpatrick instance \cite{sherrington1975,Rie94,Kis23,Koh16} with $n=10$. The exact minimum gap occurs at
\begin{equation}
s_{\rm ED}^{}=0.883,
\end{equation}
whereas the maximum of the worldline magnetization susceptibility occurs at
\begin{equation}
s_{\rm SQ}^{}=0.920.
\end{equation}
The two characteristic locations therefore differ by
\begin{equation}
\left|s_{\rm ED}^{}-s_{\rm SQ}^{}\right|=0.037,
\end{equation}
which is small relative to the complete annealing interval.

This agreement should not be interpreted as an exact identity between $\chi_m(s)$ and $\Delta(s)^{-2}$, nor as evidence that the susceptibility reconstructs the detailed gap profile. The two quantities have different definitions and arise from different descriptions of the system. The relevant observation is instead that the peak of $\chi_m(s)$ locates the same broad region in which the spectral gap becomes smallest. For schedule construction, this is the essential information: the surrogate must determine \emph{where} the anneal becomes difficult, rather than reproduce the complete instantaneous spectrum.

\begin{figure*}[t]
\centering
\includegraphics[width=0.78\linewidth]{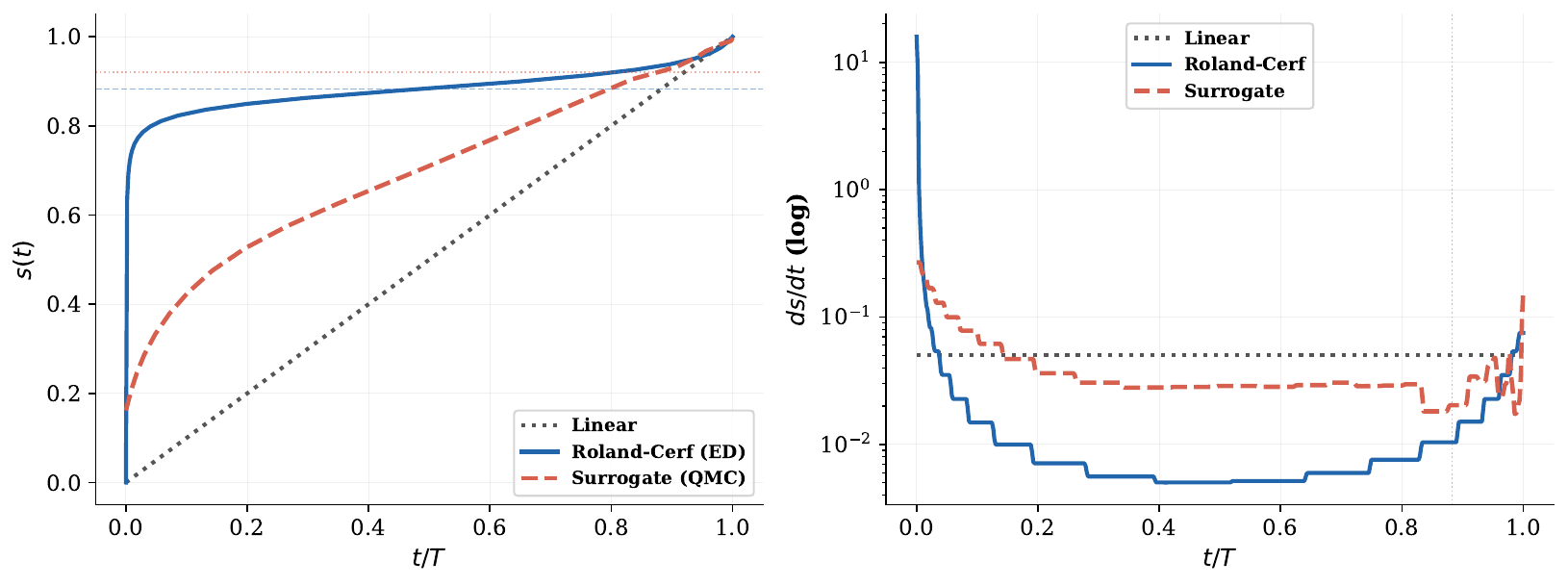}
\caption{
Comparison of the three annealing schedules for the representative $n=10$ instance at total annealing time $T=20$. Left: annealing parameter $s(t)$ as a function of normalized time $t/T$. Right: instantaneous annealing velocity $ds/dt$ on a logarithmic scale. The Roland–Cerf schedule produces an approximately four-orders-of-magnitude slowdown within a narrow interval surrounding the minimum gap, whereas the worldline-susceptibility schedule distributes the slowdown more smoothly over the broader critical region identified by $\chi_m(s)$.
}
\label{fig:schedule_comparison}
\end{figure*}

The practical consequence of this correspondence is shown in Fig.~\ref{fig:schedule_comparison}, which compares the linear, Roland–Cerf, and worldline-susceptibility schedules for the same instance at a fixed total annealing time $T=20$.

The linear schedule advances at a constant rate. The Roland–Cerf schedule responds directly to the detailed spectral-gap profile and produces an extremely sharp reduction of the annealing velocity near the minimum gap. For this instance, the local speed decreases by approximately four orders of magnitude within a narrow interval of the annealing path.

The worldline-susceptibility schedule also slows the evolution in the difficult region, but its redistribution of time is substantially broader and smoother. It therefore retains the central scheduling information supplied by the susceptibility peak without reproducing the extreme localization of the exact-gap schedule. This distinction between \emph{locating} the difficult region and \emph{following} the detailed spectral structure will become central when we compare the finite-time performance of the three schedules.

Figures~\ref{fig:gap_chiB} and \ref{fig:schedule_comparison} establish the first result of this work: an equilibrium response extracted from worldline quantum Monte Carlo contains sufficient information to identify the difficult region of the anneal and construct a nonuniform schedule without diagonalizing the quantum Hamiltonian. The next question is whether this smoother, SQA-derived schedule improves the probability of preparing the ground state at finite annealing time.

\subsection{Finite-Time Performance of the Surrogate Schedule}

Having established that the worldline magnetization susceptibility identifies the minimum-gap region and generates a smoother nonuniform schedule, we now compare the finite-time performance of the three scheduling protocols \cite{Cul11,Rak22}. For each schedule, the time-dependent Schr\"odinger equation is evolved for the same total annealing time $T$. The performance is quantified by the final ground-state probability,
\begin{equation}
P_{\rm GS}(T)
=
\left|
\left\langle
\psi_{\rm GS}
\middle|
\psi(T)
\right\rangle
\right|^2,
\label{eq:PGS}
\end{equation}
where $\ket{\psi(T)}$ is the state reached at the end of the anneal and $\ket{\psi_{\rm GS}}$ is the ground state of the final problem Hamiltonian $\hat H_P$.

\begin{figure*}[t]
\centering
\includegraphics[width=0.78\linewidth]{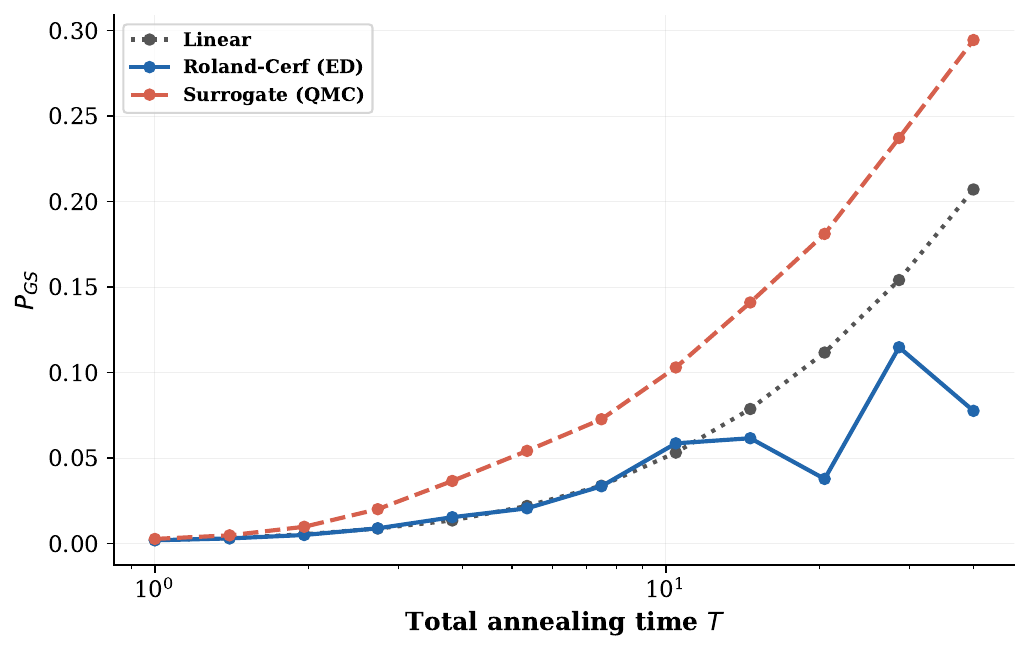}
\caption{
Final ground-state probability as a function of total annealing time for a representative $n=10$ Sherrington--Kirkpatrick instance. The worldline-susceptibility schedule consistently outperforms the linear schedule and varies smoothly with $T$. In contrast, the Roland–Cerf schedule exhibits pronounced non-monotonic behaviour, including a strong suppression near $T\approx20$, despite being constructed from the exact instantaneous spectral gap.
}
\label{fig:pgs_single_n10}
\end{figure*}

Figure~\ref{fig:pgs_single_n10} shows $P_{\rm GS}(T)$ for the representative $n=10$ instance considered in Figs.~\ref{fig:gap_chiB} and \ref{fig:schedule_comparison}. The worldline-susceptibility schedule produces the largest ground-state probability throughout the annealing-time interval investigated. Its performance increases smoothly with $T$ and remains consistently above that of the linear schedule.

The Roland–Cerf schedule behaves qualitatively differently. Despite being constructed from exact spectral information, its ground-state probability is strongly non-monotonic in the total annealing time. The success probability decreases substantially near $T\approx20$, subsequently recovers, and then decreases again at larger $T$. Thus, within the finite-time regime studied here, increasing the total runtime does not necessarily improve the performance of the exact local-adiabatic schedule.

This behaviour is not anticipated by a simple interpretation of local-adiabatic scheduling, according to which a more accurate allocation of time near the minimum-gap region should improve ground-state tracking. The result instead demonstrates that finite-time performance depends not only on whether the schedule slows down near the minimum gap, but also on how sharply the runtime is concentrated within that region.

\begin{figure*}[t]
\centering
\includegraphics[width=0.78\linewidth]{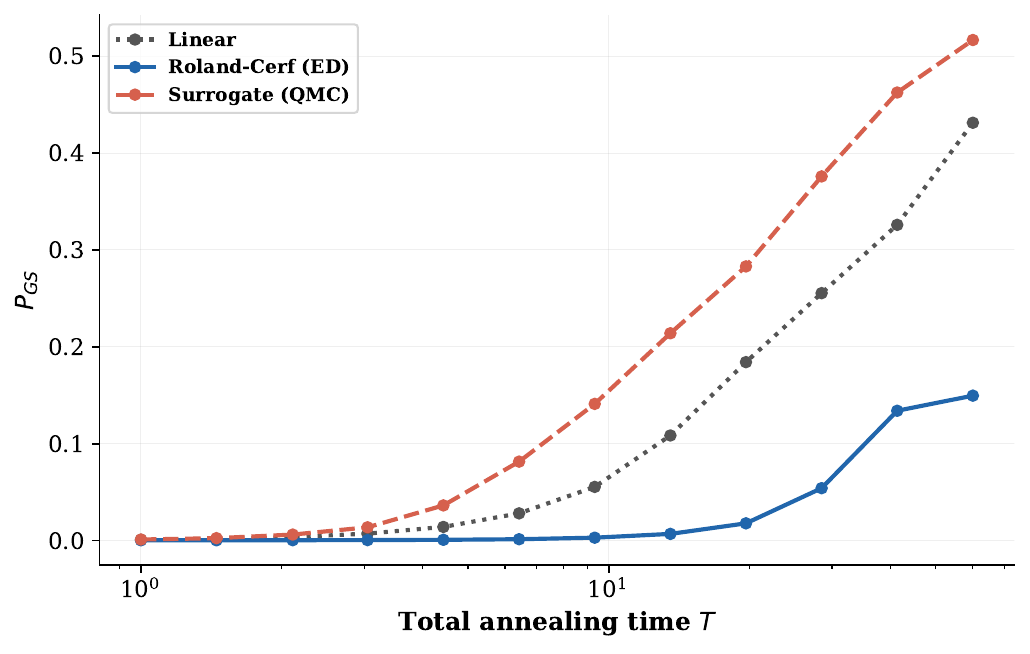}
\caption{
Final ground-state probability as a function of total annealing time for a representative $n=12$ Sherrington--Kirkpatrick instance. The exact minimum gap lies at the terminal point, $s_{\rm ED}^{}=1$, and the Roland–Cerf schedule remains strongly suppressed throughout the runtime interval. The worldline-susceptibility schedule, whose susceptibility maximum occurs at the interior point $s_{\rm SQ}^{}=0.876$, consistently produces the highest success probability.
}
\label{fig:pgs_single_n12}
\end{figure*}

A qualitatively different example is shown in Fig.~\ref{fig:pgs_single_n12} for a representative $n=12$ instance. In this case, the Roland–Cerf schedule does not exhibit the same oscillatory pattern. Instead, its ground-state probability remains systematically below those of both the linear and surrogate schedules over the complete annealing-time interval. The worldline-susceptibility schedule again yields the largest success probability, while the linear schedule gives intermediate performance.

For this instance, the exact minimum gap occurs at the terminal point,
\begin{equation}
s_{\rm ED}^{}=1,
\end{equation}
whereas the worldline susceptibility reaches its maximum at the interior point
\begin{equation}
s_{\rm SQ}^{}=0.876.
\end{equation}
This distinction anticipates the boundary-gap mechanism analysed in the following subsection. The exact local-adiabatic schedule assigns a large fraction of the available runtime to the final portion of the anneal, where the transverse-field driver has already become negligibly small. The susceptibility-derived schedule instead concentrates time in the broader crossover region where the quantum fluctuations remain active.

Table~\ref{tab:summary} summarizes the characteristic annealing locations and the final ground-state probabilities for the two representative instances. For the $n=10$ instance, the worldline-susceptibility schedule reaches
\begin{equation}
P_{\rm GS}=0.295,
\end{equation}
compared with $0.207$ for the linear schedule and $0.078$ for the Roland–Cerf schedule. For the $n=12$ instance, the corresponding probabilities are $0.517$, $0.431$, and $0.150$, respectively.

\begin{table}[htbp]
\centering
\caption{
Characteristic annealing locations and final ground-state probabilities for the two representative Sherrington–Kirkpatrick instances. The success probabilities are reported at the largest annealing time shown in Figs.~\ref{fig:pgs_single_n10} and \ref{fig:pgs_single_n12}.
}
\label{tab:summary}
\begin{tabular}{lccccc}
\toprule
System
&
$s_{\rm ED}^{}$
&
$s_{\rm SQ}^{}$
&
Linear
&
Roland–Cerf
&
Surrogate
\\
\midrule
$n=10$
&
0.883
&
0.920
&
0.207
&
0.078
&
\textbf{0.295}
\\
$n=12$
&
1.000
&
0.876
&
0.431
&
0.150
&
\textbf{0.517}
\\
\bottomrule
\end{tabular}
\end{table}

These two representative instances establish the central empirical puzzle of this work. A schedule constructed from an equilibrium observable measured in a classical worldline simulation can outperform one constructed from exact spectral information. Moreover, the degradation of the Roland–Cerf schedule appears in two qualitatively different forms: a non-monotonic dependence on the total annealing time for the $n=10$ instance and persistent suppression for the terminal-gap $n=12$ instance. We next examine the corresponding time allocation to identify the physical origins of these two finite-time behaviours.

\subsection{Two Finite-Time Failure Modes of Exact Local-Adiabatic Scheduling}
\label{sec:diagnosis}

The results of the previous subsection reveal an unexpected feature of finite-time quantum annealing. The Roland--Cerf schedule is constructed using complete knowledge of the instantaneous spectral gap and is theoretically optimal within the local-adiabatic approximation. Nevertheless, for both representative instances it is consistently outperformed by the surrogate schedule constructed solely from the worldline susceptibility measured during simulated quantum annealing.

This observation suggests that, for finite annealing times, the quality of a schedule is determined not only by how accurately it follows the instantaneous spectral gap, but also by how the available runtime is distributed along the annealing trajectory. While the local-adiabatic prescription is derived to minimize diabatic transitions in the asymptotic adiabatic limit, practical quantum annealing necessarily operates at finite annealing times, where the detailed distribution of runtime can itself influence the dynamics \cite{Gar22,Mut15}.

Our numerical results reveal two qualitatively distinct mechanisms by which an exact implementation of the Roland--Cerf schedule can become counterproductive.

The first occurs when the minimum spectral gap is located at, or extremely close to, the end of the annealing path. In this case, the local-adiabatic prescription allocates a disproportionately large fraction of the total runtime to the final stage of the evolution, where the transverse-field driver has already become negligibly small. We refer to this mechanism as the \emph{boundary-gap trap} \cite{Jor09,You09}.

The second mechanism arises when the minimum gap is located well inside the annealing interval. Here the local-adiabatic schedule concentrates an overwhelming fraction of the available runtime into an extremely narrow neighbourhood of the avoided crossing. Rather than improving the evolution, this highly localized slowdown generates pronounced oscillations in the final ground-state probability as the total annealing time is varied. As we demonstrate below, these oscillations cannot be explained by conventional two-level Landau--Zener--Stückelberg interference and instead indicate a genuinely multilevel finite-time effect.

Both mechanisms can be understood by examining the cumulative time-allocation function,
\begin{equation}
F(s)=
\frac{\displaystyle\int_0^s \Delta(s')^{-2}\,ds'}
{\displaystyle\int_0^1 \Delta(s')^{-2}\,ds'},
\label{eq:Fcum}
\end{equation}
which measures the fraction of the total annealing time allocated before reaching annealing parameter $s$. Regions where $F(s)$ rises rapidly correspond to intervals in which a disproportionately large fraction of the total runtime is concentrated.

\begin{figure*}[t]
\centering
\includegraphics[width=0.78\linewidth]{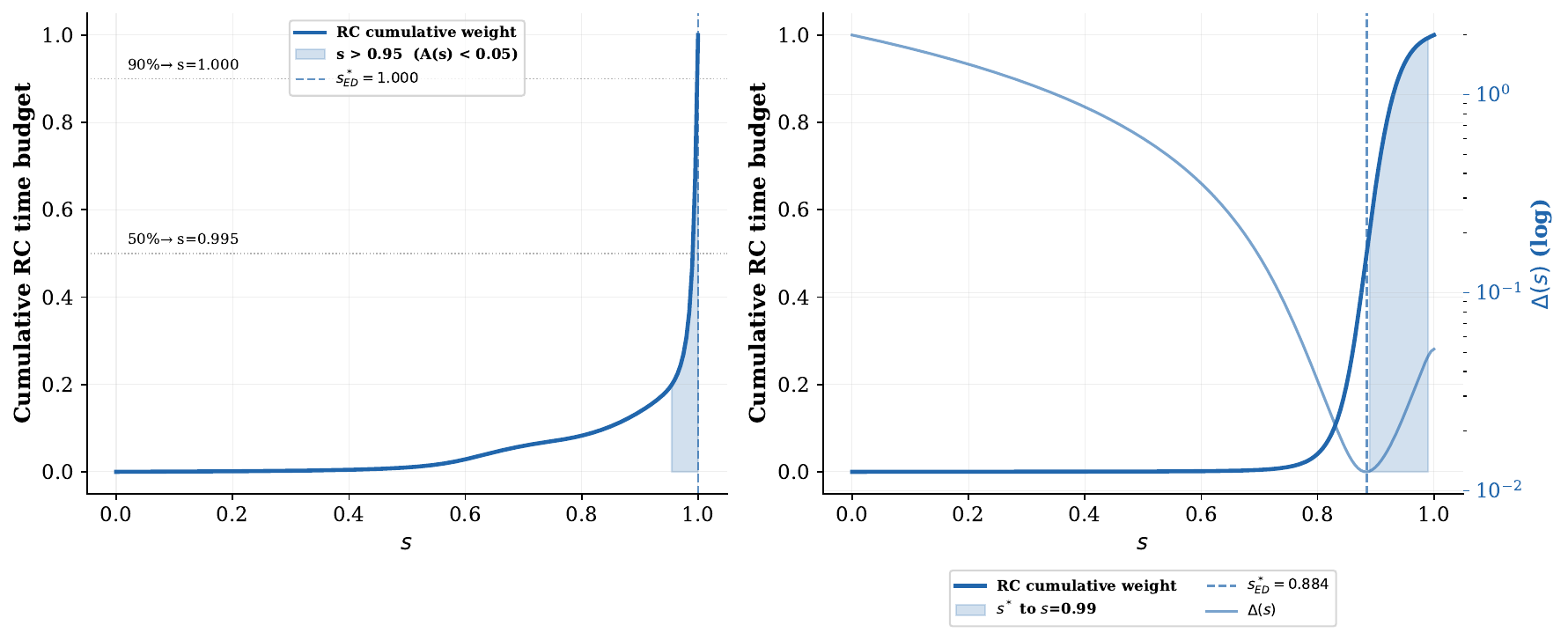}
\caption{
Cumulative time allocation for the Roland--Cerf schedule. The function $F(s)$ gives the fraction of the total annealing time spent before reaching annealing parameter $s$. Left: representative $n=12$ instance exhibiting the boundary-gap trap, where most of the runtime is concentrated near the endpoint $s=1$. Right: representative $n=10$ instance, where the runtime is concentrated into a narrow interval surrounding the interior minimum gap. These two time-allocation patterns give rise to two distinct finite-time failure modes of exact local-adiabatic scheduling.
}
\label{fig:rc_failure_mechanisms}
\end{figure*}

Figure~\ref{fig:rc_failure_mechanisms} illustrates these two representative cases. For the $n=12$ instance, nearly the entire runtime accumulates near the endpoint of the anneal, producing the boundary-gap trap. For the $n=10$ instance, the runtime is instead compressed into a narrow interval surrounding the interior minimum gap. Although both schedules satisfy exactly the same local-adiabatic condition, they produce two qualitatively different finite-time dynamical pathologies.

The physical origin of these two mechanisms is analysed separately in the following subsections.

\subsubsection{Boundary-Gap Trap}

The first finite-time failure mode occurs when the minimum spectral gap is located at, or extremely close to, the end of the annealing trajectory. In such instances, the Roland--Cerf prescription faithfully allocates the majority of the available runtime to the region where the gap is smallest. While this allocation is formally consistent with the local-adiabatic condition, it becomes dynamically ineffective because the minimum gap appears only after the transverse-field driver has almost completely vanished.

This behaviour is illustrated by the representative $n=12$ instance shown in the left panel of Fig.~\ref{fig:rc_failure_mechanisms}. The exact minimum gap occurs at

\[
s_{\rm ED}^{*}=1.000,
\]

causing the cumulative time-allocation function to increase sharply only at the very end of the anneal. Approximately one-half of the total annealing time is spent after

\[
s\approx0.995,
\]

while nearly ninety percent of the runtime is concentrated within the final fraction of the annealing trajectory.

At first sight this behaviour appears optimal because it follows directly from the exact spectral information. However, the physical dynamics near the endpoint differs fundamentally from that near an interior avoided crossing. As

\[
s\rightarrow1,
\]

the system Hamiltonian approaches

\[
H(s)\approx H_P,
\]

while the transverse-field driver satisfies

\[
(1-s)H_D\rightarrow0.
\]

Consequently, the Hamiltonian has already become almost entirely classical. The transverse field, which provides the quantum fluctuations responsible for tunnelling between computational basis states, has effectively disappeared. Additional evolution time in this regime therefore contributes little to improving the probability of reaching the ground state, even though the instantaneous spectral gap is smallest.

The poor performance of the Roland--Cerf schedule in this case should therefore not be interpreted as a failure of the local-adiabatic principle itself. Rather, it reflects a limitation of using the instantaneous spectral gap as the sole indicator for allocating computational effort during finite-time evolution. When the minimum gap is pushed to the boundary of the annealing path, the gap no longer coincides with the region where quantum fluctuations are most effective. The schedule therefore invests most of its runtime in a region where the dynamics has already become largely frozen.

The worldline-susceptibility schedule naturally avoids this pathology. Instead of tracking the exact minimum gap, it follows the fluctuations of the worldline magnetization measured during the SQA simulation. For the same instance, the susceptibility reaches its maximum at

\[
s_{\rm SQ}^{*}=0.876,
\]

well before the endpoint of the anneal. The resulting schedule therefore slows the evolution throughout the broader crossover region where the transverse field remains appreciable and quantum fluctuations are still active. Rather than concentrating nearly the entire runtime at the endpoint, it distributes the available annealing time over the physically relevant part of the evolution.

This behaviour explains the substantial improvement in the final ground-state probability observed previously in Fig.~\ref{fig:pgs_single_n12}. More generally, it suggests that, for finite-time quantum annealing, the most useful scheduling indicator is not necessarily the exact location of the minimum spectral gap, but the broader crossover region where quantum fluctuations remain sufficiently strong to influence the dynamics.

\subsubsection{Oscillatory Instability}

The second finite-time failure mode is qualitatively different from the boundary-gap trap. Here the minimum spectral gap is located well inside the annealing interval rather than at its endpoint. One would therefore expect the Roland--Cerf schedule to provide an excellent approximation to the optimal adiabatic evolution. Surprisingly, the representative $n=10$ instance exhibits precisely the opposite behaviour.

As shown previously in Fig.~\ref{fig:pgs_single_n10}, the final ground-state probability does not increase monotonically with the total annealing time. Instead, the success probability repeatedly increases and decreases as the runtime is varied. Representative values are listed in Table~\ref{tab:stuckelberg}. Around
\[
T\approx20,
\]
the Roland--Cerf schedule reaches one of its deepest minima, while both the linear schedule and the worldline-susceptibility schedule continue to improve smoothly.

\begin{table}[htbp]
\centering
\caption{Representative ground-state probabilities for the $n=10$ instance. The worldline-susceptibility schedule improves monotonically, whereas the Roland--Cerf schedule exhibits pronounced oscillatory behaviour as the total annealing time is varied.}
\label{tab:stuckelberg}
\begin{tabular}{lccc}
\toprule
$T$ & Linear & Roland--Cerf & Surrogate\\
\midrule
10.46 & 0.053 & 0.059 & 0.103\\
14.63 & 0.079 & 0.062 & 0.141\\
20.45 & 0.112 & 0.038 & 0.181\\
28.60 & 0.154 & 0.115 & 0.237\\
40.00 & 0.207 & 0.078 & 0.295\\
\bottomrule
\end{tabular}
\end{table}

\begin{figure*}[t]
\centering
\includegraphics[width=0.78\linewidth]{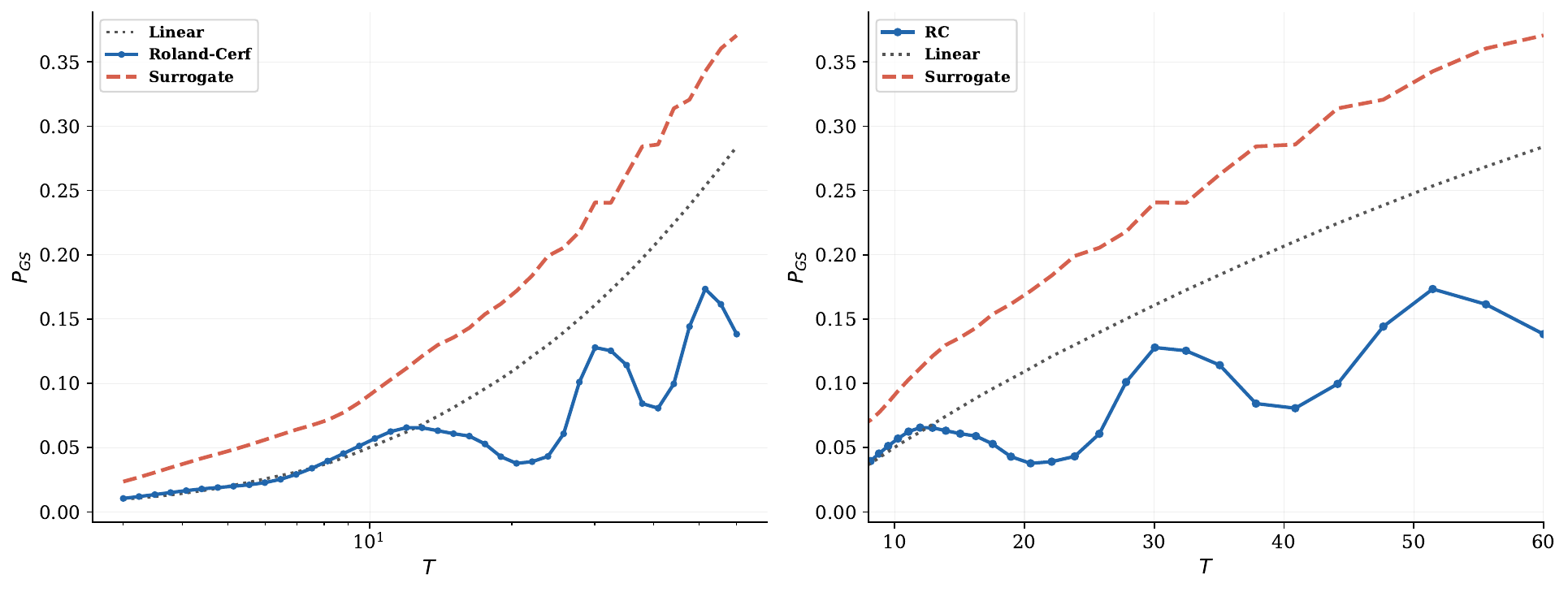}
\caption{
Dense scan of the total annealing time for the representative $n=10$ Sherrington--Kirkpatrick instance. The Roland--Cerf schedule exhibits a reproducible oscillatory dependence of the final ground-state probability, whereas both the linear and worldline-susceptibility schedules remain smooth and monotonic.
}
\label{fig:dense_T}
\end{figure*}

To determine whether this behaviour represents an isolated numerical fluctuation or a genuine dynamical phenomenon, we computed the ground-state probability over a dense grid of annealing times. The resulting evolution is shown in Fig.~\ref{fig:dense_T}. Rather than random fluctuations, the data reveal a reproducible oscillatory pattern extending across the interval

\[
3\le T\le60,
\]

with approximately five well-defined extrema. In contrast, both the linear schedule and the worldline-susceptibility schedule evolve smoothly over the same interval.

Since such oscillations could potentially arise from numerical inaccuracies, we repeated the calculations using independent numerical integrators, different schedule discretizations, and integration tolerances ranging from $10^{-8}$ to $10^{-12}$. In every case the oscillatory pattern remained unchanged within numerical precision, demonstrating that it is not a solver-dependent artifact.

A natural explanation is coherent Landau--Zener--Stückelberg (LZS) interference within an effective two-level system \cite{shevchenko2010}. To test this possibility, we estimated the oscillation period predicted from the accumulated dynamical phase. The predicted period is

\[
T_{\rm LZS}\approx312,
\]

whereas the observed oscillation period is only

\[
T_{\rm obs}\approx20.
\]

The discrepancy exceeds an order of magnitude, indicating that the conventional two-level LZS mechanism cannot account for the observed oscillations.

We therefore attribute this behaviour to a multilevel finite-time instability generated by the extreme localization of the Roland--Cerf schedule. As shown in the right panel of Fig.~\ref{fig:rc_failure_mechanisms}, the exact local-adiabatic prescription concentrates a large fraction of the available runtime into a very narrow interval surrounding the minimum spectral gap. While this suppresses transitions between the ground and first excited states, it also amplifies coherent interference involving higher excited states. The resulting multilevel interference produces the oscillatory dependence of the final ground-state probability.

The worldline-susceptibility schedule naturally avoids this instability because it distributes the slowdown over a substantially broader region of the annealing path. Rather than reproducing every detail of the spectral-gap profile, it captures the location of the critical region while avoiding excessive localization of the runtime. Consequently, the evolution remains considerably more stable and the ground-state probability increases smoothly with the total annealing time.

The oscillatory instability therefore represents a second finite-time limitation of exact local-adiabatic scheduling. Unlike the boundary-gap trap, which originates from the location of the minimum spectral gap, this mechanism arises from an overly aggressive concentration of runtime around an interior minimum. Together, these two mechanisms explain why a schedule constructed from an inexpensive equilibrium observable measured in simulated quantum annealing can outperform one constructed from complete spectral information under realistic finite-time conditions.

\subsection{Statistical Validation Across Disorder Realizations and System Sizes}

The representative examples discussed above establish the existence of two distinct finite-time failure modes of the exact local-adiabatic schedule. An important question, however, is whether these mechanisms are exceptional or whether they persist across independent disorder realizations and increasing system size.

To address this question, we carried out large-scale numerical experiments using the \texttt{Qanneal} framework developed for this work. The framework automates the complete computational pipeline, including simulated quantum annealing, measurement of the worldline magnetization susceptibility, construction of surrogate schedules, exact spectral calculations where feasible, real-time Schr\"odinger evolution, and statistical analysis over ensembles of disorder realizations. This automated workflow makes it possible to compare scheduling strategies systematically across many independent instances and multiple system sizes.

\begin{figure}[htbp]
\centering
\includegraphics[width=1.0\linewidth]{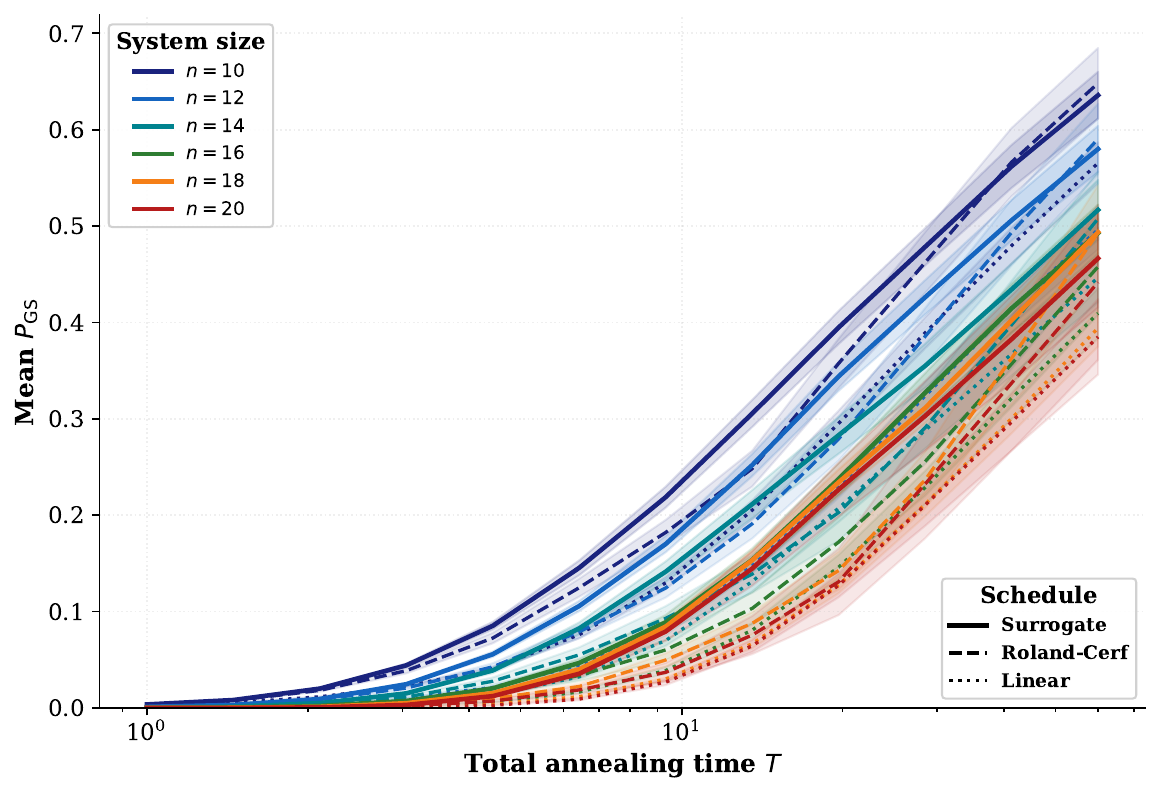}
\caption{
Disorder-averaged ground-state probability for Sherrington--Kirkpatrick instances with
$n=\{10,12,14,16,18,20\}$.
Each panel shows the mean over independent disorder realizations, with shaded regions denoting $\pm1$ standard error of the mean. Across all system sizes, the worldline-susceptibility schedule consistently achieves the highest average success probability. The annotations indicate the numbers of instances classified as boundary-gap (B), oscillatory (O), and conventional (C).
}
\label{fig:pgs_mean_all_n}
\end{figure}

Figure~\ref{fig:pgs_mean_all_n} summarizes the disorder-averaged ground-state probability for Sherrington--Kirkpatrick spin-glass instances with

\[
n\in\{10,12,14,16,18,20\},
\]

where each panel shows the mean performance over independent disorder realizations together with the corresponding standard error of the mean.

Three observations immediately emerge.

First, the worldline-susceptibility schedule consistently achieves the highest average ground-state probability across the entire annealing-time range for every system size considered. Although the quantitative improvement varies between disorder ensembles, the qualitative ordering of the three schedules remains unchanged.

Second, the advantage of the surrogate schedule is not restricted to the smallest systems for which exact diagonalization is straightforward. The same trend persists as the Hilbert-space dimension increases, demonstrating that the proposed construction does not rely on accidental features of a few representative instances.

Third, the Roland--Cerf schedule does not uniformly outperform the linear schedule despite having access to complete spectral information. As the system size increases, the fraction of instances exhibiting the boundary-gap trap also increases, leading to a progressive degradation of the disorder-averaged performance of the exact local-adiabatic schedule relative to the smoother worldline-susceptibility schedule.

\begin{figure}[ht]
\centering
\includegraphics[width=1.0\linewidth]{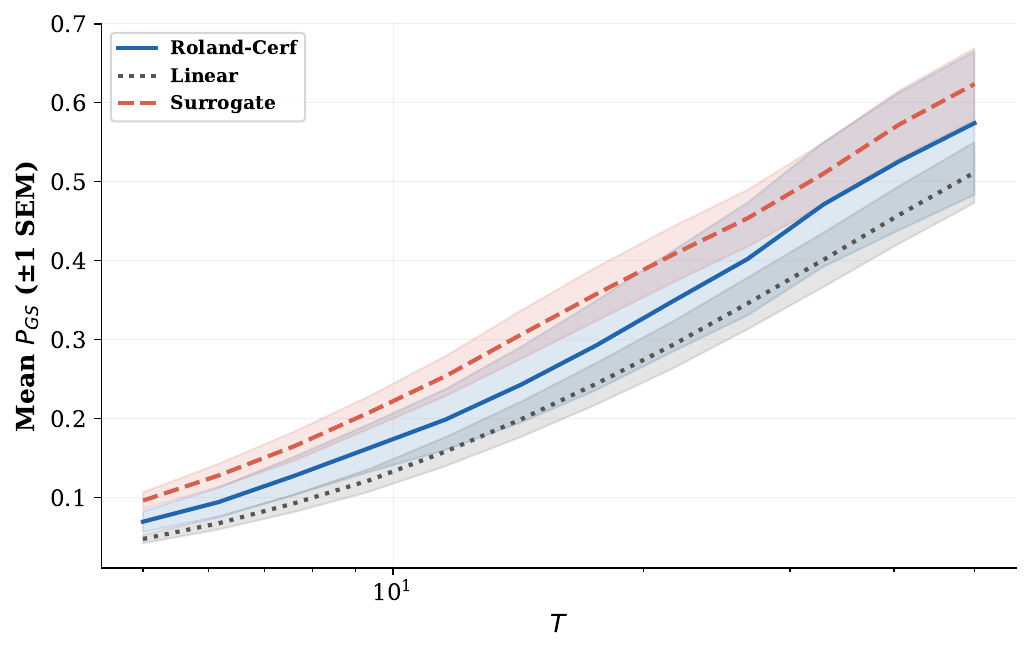}
\caption{
Average ground-state probability for twenty independent $n=10$ Sherrington--Kirkpatrick instances. The worldline-susceptibility schedule consistently provides the highest average success probability throughout the annealing interval. The labels B, O, and C denote the numbers of boundary-gap, oscillatory, and conventional instances within the ensemble.
}
\label{fig:pgs_mean_n10}
\end{figure}

\begin{figure}[htbp]
\centering
\includegraphics[width=1.0\linewidth]{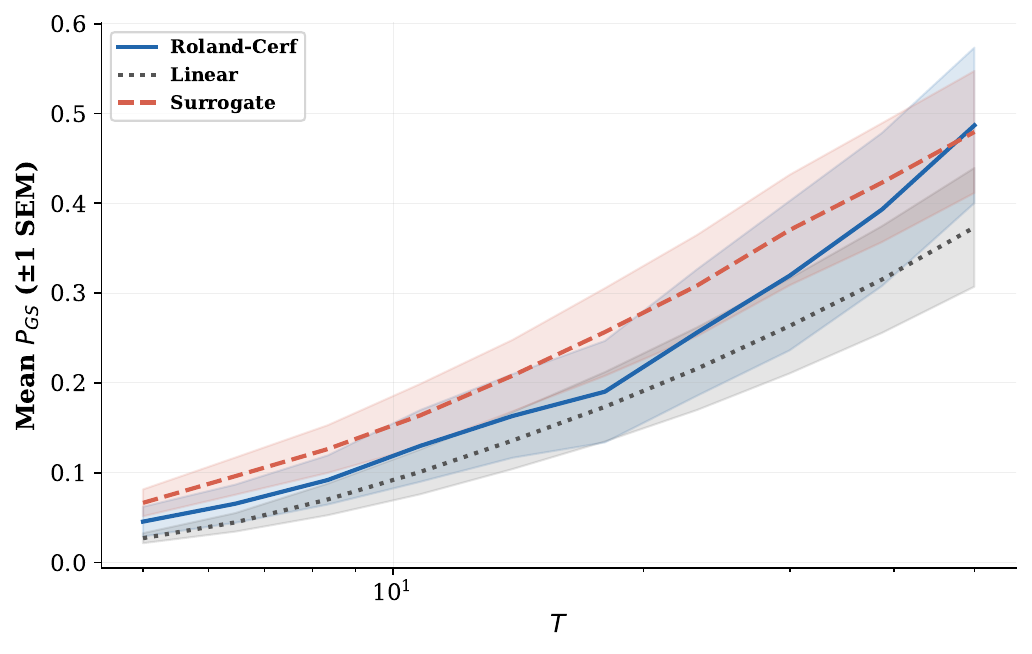}
\caption{
Average ground-state probability for ten independent $n=12$ Sherrington--Kirkpatrick instances. The increased prevalence of boundary-gap instances leads to a stronger suppression of the Roland--Cerf schedule, whereas the worldline-susceptibility schedule maintains the highest average success probability throughout the annealing-time range.
}
\label{fig:pgs_mean_n12}
\end{figure}

The smallest system sizes, for which the largest numbers of independent disorder realizations were analysed, are shown separately in Figs.~\ref{fig:pgs_mean_n10} and \ref{fig:pgs_mean_n12}. These enlarged views demonstrate that the improvement observed for the representative examples survives disorder averaging and therefore reflects generic behaviour of the ensemble rather than isolated exceptional instances.

Taken together, these results demonstrate that the superiority of the worldline-susceptibility schedule is not confined to a few representative examples. Instead, it persists after averaging over disorder realizations and remains robust across all system sizes accessible to exact benchmarking in the present work. The representative instances analysed in the previous subsections therefore illustrate generic finite-time mechanisms rather than isolated anomalies.

\section{Conclusion}

We have introduced a practical surrogate for local-adiabatic quantum annealing based on the worldline magnetization susceptibility measured during simulated quantum annealing. Unlike the Roland--Cerf schedule, which requires complete knowledge of the instantaneous spectral gap and repeated diagonalization of the quantum Hamiltonian, the proposed approach constructs the annealing schedule entirely from equilibrium observables obtained during an SQA simulation. The resulting procedure therefore remains computationally inexpensive and naturally extends to problem sizes for which exact spectral calculations are infeasible.

Using exact diagonalization as ground truth for representative Sherrington--Kirkpatrick spin-glass instances, we demonstrated that the worldline susceptibility reliably identifies the minimum-gap region of the annealing process. The resulting schedules consistently outperform conventional linear annealing and, remarkably, also outperform the exact Roland--Cerf schedule for a substantial fraction of the instances investigated.

To explain this seemingly counterintuitive observation, we identified two distinct finite-time failure modes of exact local-adiabatic scheduling. The first is a boundary-gap trap, in which the minimum spectral gap occurs at the end of the annealing path, causing the schedule to allocate most of the available runtime after the transverse-field driver has effectively disappeared. The second is an oscillatory instability arising from an excessively localized distribution of annealing time around an interior minimum gap, producing coherent multilevel interference during finite-time evolution. Together, these mechanisms demonstrate that exact knowledge of the instantaneous spectral gap does not necessarily lead to the most effective schedule under realistic finite-time conditions.

The proposed surrogate avoids these pathologies by identifying the broader crossover region where quantum fluctuations remain dynamically relevant rather than reproducing the detailed structure of the instantaneous spectrum. Consequently, the schedule distributes the available runtime more smoothly while preserving the essential information required for efficient ground-state preparation.

The methodology developed here has been implemented in the open-source \texttt{Qanneal} framework, which integrates simulated quantum annealing, worldline susceptibility measurements, surrogate schedule construction, exact spectral benchmarking, and quantum-dynamical simulations within a unified computational platform. Beyond the present study of Sherrington--Kirkpatrick spin glasses, this framework provides a practical environment for investigating scheduling strategies for general transverse-field Ising optimization problems \cite{Fin23,Mor07c}.

More broadly, our results suggest a shift in perspective for quantum annealing schedule design. Rather than seeking increasingly accurate reconstructions of the instantaneous spectral gap, it may be more advantageous to construct schedules from inexpensive equilibrium observables that robustly identify the physically relevant region of the annealing process. This observation opens a scalable route toward schedule optimization that combines classical Monte Carlo simulations with quantum annealing, without requiring explicit access to the exponentially expensive quantum spectrum \cite{Hei14,Kin22b}.

\begin{acknowledgments}
The authors acknowledge support from the National Quantum Mission (NQM), an initiative of the Department of Science and Technology (DST), Government of India under the Project titled \emph{Q-LAT Anneal A - General-Purpose QUBO Compiler for Quantum Many-Body Physics and
Hybrid Optimisation}.
S.S. and L.N. were supported by the DAE Research Visitor Fellowship at The Institute of Mathematical Sciences (IMSc), Chennai. We also thank IMSc for providing HPC resources.
\end{acknowledgments}
\newpage
\bibliographystyle{apsrev4-2}
\bibliography{references}
\clearpage
\appendix

\section{Qanneal: An Open-Source Framework for Simulated Quantum Annealing}
\label{app:qanneal}
All simulated-quantum-annealing results in this work were produced with
\texttt{qanneal}, an open-source, research-grade Ising/QUBO
annealing library written in C++17 with a Python front end (pybind11
bindings), released under the Apache License 2.0 \cite{Wil21}. The package implements
five annealing engines behind a single high-level entry point,
\texttt{solve()}, and two theoretically grounded adaptive-schedule
constructions built on a measured susceptibility as a proxy for the
(unknown) spectral gap. This appendix summarizes installation, the
theoretical machinery, and the exact commands needed to reproduce the
simulations reported in the main text. The complete derivations, full
API reference, and additional worked examples are given in the package
manual~\cite{qanneal_manual}.

\subsection{Installation}

\texttt{qanneal} requires Python~$\geq 3.11$, NumPy, and a C++17
compiler (GCC~$\geq 9$, Clang~$\geq 10$, or MSVC~2019+); \texttt{OpenMP}
is optional but recommended for replica/slice parallelism, and
\texttt{dimod}/\texttt{networkx} are optional for BQM/graph problem
inputs. From the repository root:

\begin{lstlisting}[language=bash]
python -m pip install . --no-build-isolation

python -m pip install -e . --no-build-isolation

python -c "import qanneal; print(qanneal.__version__)"
\end{lstlisting}

The build compiles the C++17 core and the \texttt{qanneal.\_qanneal}
pybind11 extension in place; no external references or additional
dependencies are needed to reproduce any result in this paper. On
macOS, Apple's Clang ships without \texttt{OpenMP}; install it with
\texttt{brew install libomp} before building, or the package will run
correctly but single-threaded. For C++-only use or the test suite:

\begin{lstlisting}[language=bash]
cmake -S . -B build -DQANNEAL_ENABLE_OPENMP=ON \
                     -DCMAKE_BUILD_TYPE=Release
cmake --build build -j
ctest --test-dir build
\end{lstlisting}

\subsection{Problem setup}

Both the Ising and QUBO conventions of Eqs.~(1)--(2) of the main text
are accepted directly; the package auto-detects dense arrays, sparse
edge lists, dictionaries, \texttt{dimod} binary quadratic models, and
\texttt{networkx} graphs. A Sherrington--Kirkpatrick instance, as used
throughout this paper, is built as a dense, fully-connected Ising model:

\begin{lstlisting}[language=Python]
import numpy as np
from qanneal import DenseIsing

rng = np.random.default_rng(seed)
n = 10
J = np.triu(rng.standard_normal((n, n)), 1)
J = (J + J.T) / np.sqrt(n)
h = np.zeros(n)
ising = DenseIsing(h, J)
\end{lstlisting}

\subsection{Suzuki--Trotter representation}

\texttt{qanneal} samples the equilibrium density matrix of the
transverse-field Ising Hamiltonian, Eq.~(1) of the main text, by
Trotterizing the imaginary-time partition function into $M$ classical
replicas (``slices'') coupled along imaginary time with strength
\begin{equation}
J_\perp(\beta,\Gamma,M) = \frac{1}{2}\ln\coth\!\left(\frac{\beta\Gamma}{M}\right),
\end{equation}
giving the effective classical action of Eq.~(5) of the main text,
\begin{equation}
S_{\rm eff} = \frac{\beta}{M}\sum_{k=1}^{M} H_P\!\left(s^{(k)}\right)
  - J_\perp \sum_{k=1}^{M}\sum_{i=1}^{N} s_i^{(k)} s_i^{(k+1)},
\end{equation}
with periodic boundary conditions in imaginary time, $s^{(M+1)}\equiv
s^{(1)}$. Monte Carlo updates combine (i) single-spin (slice) moves,
whose acceptance follows the ordinary Metropolis rule applied to
$\Delta S$; (ii) whole-worldline flips, for which the imaginary-time
coupling cancels identically, making them essential once $J_\perp$ is
large; and (iii) Swendsen--Wang cluster moves \cite{swendsen1987} along the imaginary-time
direction, which join aligned time-bonds with probability
$1-e^{-2J_\perp}$ and interpolate between the two extremes. All three
satisfy detailed balance for $S_{\rm eff}$; full derivations of the
acceptance rules are given in Secs.~5.2--5.3 of the manual \cite{Mar18,Alb20}.

\subsection{Engines available}

\begin{table}[h]
\centering
\small
\begin{tabular}{@{}p{0.32\linewidth}p{0.58\linewidth}@{}}
\hline\hline
Engine (\texttt{method=}) & What it simulates \\
\hline
\texttt{"sa"} & Classical simulated annealing (Metropolis) \\
\texttt{"sqa"} & Discrete-time Suzuki--Trotter SQA \\
\texttt{"sqapt"} & SQA + replica exchange, $(\beta,\Gamma)$ ladder \\
\texttt{"ctpimc"} & Continuous-time path-integral Monte Carlo \cite{martonak2002} \\
\texttt{"sqa\_chi"} & SQA with the worldline-susceptibility schedule \\
\hline\hline
\end{tabular}
\caption{Annealing engines provided by \texttt{qanneal} through the
common \texttt{solve()} interface. The surrogate schedule of this paper
is run through \texttt{sqa\_chi} (Appendix~\ref{app:sqachi}); the
linear and Roland--Cerf comparisons use \texttt{SQAAnnealer} to measure
$\chi_m(s)$ and standard sampling, with the two schedules' trajectories
supplied externally, since both are precomputed from a prescribed
$s(t)$ rather than constructed by the package.}
\label{tab:engines}
\end{table}

A single call runs a standard schedule end to end:
\begin{lstlisting}[language=Python]
from qanneal import solve
result = solve(ising, method="sqa", reads=16,
               trotter_slices=32, replicas=4,
               sweeps_per_beta=60, worldline_sweeps=4,
               seed=0)
print(result.best_energy, result.best_sample)
\end{lstlisting}

\subsection{The worldline-susceptibility schedule (\texttt{sqa\_chi})}
\label{app:sqachi}

The surrogate schedule constructed in Sec.~II.C of the main text is
implemented directly as the \texttt{SQAChiAnnealer} class. Its
execution mirrors the workflow of Sec.~II.D one-to-one:

\begin{enumerate}
\item \emph{Pilot scan.} $\chi_m(s)$ [Eq.~(8) of the main text] is
measured at \texttt{scan\_points} values of the annealing parameter
$s\in[0,1]$, each from an independently randomized worldline (not
carried over between grid points, so every estimate is statistically
independent), using \texttt{scan\_burn} burn-in sweeps followed by
\texttt{scan\_sweeps} measurement sweeps of a parity-parallel
checkerboard kernel: since Trotter slices of the same parity are never
directly coupled, even and odd slices are each updated in one
\texttt{OpenMP} parallel pass, giving exact detailed balance for any
coupling graph while still parallelizing within a single replica.
\item \emph{Time allocation.} The pilot profile is floor-regularized,
\begin{equation}
w(s) = \max\!\big(\chi_m(s),\ \texttt{chi\_floor\_fraction}\cdot
\max_{s'}\chi_m(s')\big),
\end{equation}
and inverted via the cumulative map of Eqs.~(12)--(14) of the main text
to place \texttt{num\_steps} schedule points at equal increments of
accumulated susceptibility weight, precomputing the full trajectory
rather than integrating it online (avoiding compounding-noise
failure).
\item \emph{Production anneal.} A short thermal ramp (fraction
\texttt{beta\_ramp\_fraction} of the step budget, default $0.3$)
equilibrates the worldlines at the target inverse temperature at fixed
$\gamma_{\rm start}$; the $\gamma(s)$ trajectory from step 2 is then
executed on a fresh worldline through the same checkerboard kernel.
\end{enumerate}

\begin{lstlisting}[language=Python]
from qanneal import SQAChiAnnealer

ann = SQAChiAnnealer(ising, trotter_slices=M, replicas=4)
ann.set_seed(seed)
result = ann.run_chi(
    beta=beta,
    gamma_start=gamma_start, gamma_end=gamma_end,
    num_steps=200, sweeps_per_step=20,
    scan_points=16, scan_sweeps=30, scan_burn=10,
    chi_floor_fraction=1e-6,
    beta_ramp_fraction=0.3,
)

s_star     = result.s_star
scan_s     = result.scan_s
scan_chiB  = result.scan_chi_B
gamma_sch  = result.gamma_schedule
s_sch      = result.s_schedule
\end{lstlisting}

Both the pilot scan and the production run report their full
trajectories for auditing (\texttt{scan\_s}, \texttt{scan\_chi\_B},
\texttt{gamma\_schedule}, \texttt{s\_schedule}), and the estimated
critical point $s^{*}_{\rm SQ}$ used throughout Sec.~III is returned
directly as the location of the $\chi_m(s)$ peak on the pilot grid.
The resolved \texttt{s\_schedule} is the trajectory $s(t)$ passed to
the external time-dependent Schr\"odinger propagation used to obtain
the ground-state probabilities of Figs.~3--13; the linear and
Roland--Cerf trajectories used for comparison are constructed directly
from $s(t)=t/T$ and from Eq.~(4) of the main text respectively, using
the same propagator.

\subsection{Reproducing the results of this paper}

The parameter grid of Sec.~III.G (Figs.~14--15) is reproduced by
looping \texttt{run\_chi} over
\begin{align}
M &\in \{8, 16, 32, 64\}, \\
\beta &\in \{2, 5, 10, 20\},
\end{align}
with \texttt{scan\_points=16}, \texttt{scan\_sweeps=30},
\texttt{scan\_burn=10}, \texttt{replicas=4}, and comparing the
resulting \texttt{s\_star} against the exact-diagonalization estimate
$s^{*}_{\rm ED}$ for the same instance. A minimal reproduction script:

\begin{lstlisting}[language=Python]
import numpy as np
from qanneal import DenseIsing, SQAChiAnnealer

def make_sk(n, seed):
    rng = np.random.default_rng(seed)
    J = np.triu(rng.standard_normal((n, n)), 1)
    J = (J + J.T) / np.sqrt(n)
    return DenseIsing(np.zeros(n), J)

for n in [10, 12, 14, 16, 18, 20]:
    for seed in range(20):
        ising = make_sk(n, seed)
        ann = SQAChiAnnealer(ising, trotter_slices=32,
                              replicas=4)
        ann.set_seed(seed)
        res = ann.run_chi(beta=5.0, gamma_start=5.0,
                           gamma_end=0.01, num_steps=200,
                           sweeps_per_step=20,
                           scan_points=16, scan_sweeps=30,
                           scan_burn=10,
                           chi_floor_fraction=1e-6)
\end{lstlisting}

Ground-state probabilities [Eq.~(19) of the main text] are obtained by
propagating the time-dependent Schr\"odinger equation (or the Lindblad
master equation, Eq.~(21), for the decoherence study of Sec.~III.F)
under $H(s(t))$ with $s(t)$ taken from \texttt{s\_schedule}
(surrogate), a linear ramp (linear), or the Roland--Cerf integral,
Eq.~(4), evaluated from exact-diagonalization gap data (Roland--Cerf);
this propagation step is external to \texttt{qanneal}, which supplies
only the schedule construction and the classical Monte Carlo
measurement of $\chi_m(s)$.

\subsection{Parameters used in this work}

\begin{table}[H]
\centering
\small
\begin{tabular}{@{}p{0.55\linewidth}p{0.35\linewidth}@{}}
\hline\hline
Parameter & Values \\
\hline
Trotter slices $M$ & $\{8,16,32,64\}$ \\
Inverse temperature $\beta$ & $\{2,5,10,20\}$ \\
Pilot scan points & 16 \\
Pilot scan sweeps / burn-in & 30 / 10 \\
Replicas & 4 \\
Regularization floor & $10^{-6}\times\max_s \chi_m$ \\
Thermal ramp fraction & 0.3 \\
\hline\hline
\end{tabular}
\caption{Parameter grid used for the SQA-parameter robustness study
(Sec.~III.G), all accessible through
\texttt{SQAChiAnnealer.run\_chi} or the equivalent
\texttt{solve(..., method="sqa\_chi")} interface.}
\label{tab:params}
\end{table}

\subsection{Availability}

\texttt{qanneal} is released under the Apache License 2.0. Source
code, build instructions, and the complete manual from which this
appendix is condensed are available at https://pypi.org/project/qanneal/ Users of the package for scheduling or SQA simulation are
asked to cite this paper.

\section{Supplementary Robustness Checks}
\label{app:supplementary}

The main text establishes the boundary-gap trap and the oscillatory
instability (Sec.~\ref{sec:diagnosis}) as the two dominant finite-time
failure modes of the exact Roland--Cerf schedule, and shows that both
persist under disorder averaging (Sec.~III.D). This appendix collects
five supplementary checks that were run 
to stress-test those conclusions against three concerns that naturally arises: (i) whether the oscillatory instability is a
genuine coherent effect or a fragile numerical artifact that would
wash out under any amount of environmental noise; (ii) whether the
qualitative advantage of the surrogate schedule survives the same kind
of decoherence; (iii) whether the residual disagreement between the
exact-diagonalization crossing point $s^{*}_{\rm ED}$ and the
SQA-derived crossing point $s^{*}_{\rm SQ}$ is a controllable
discretization bias of the Suzuki--Trotter representation or an
irreducible finite-size/disorder floor; and (iv)--(v) whether the
$\Delta^{*}$-based failure threshold and the disorder-averaged
failure-class rates quoted in the main text survive a held-out
statistical test and hold up at larger $n$. All calculations in this
appendix were produced with the same \texttt{qanneal} engines and
propagators described above (Appendix~\ref{app:qanneal}), applied
either to the representative $n=10$ instance of
Figs.~\ref{fig:pgs_single_n10}--\ref{fig:dense_T} or, for the
threshold and base-rate checks, to the full disorder ensemble
underlying Fig.~\ref{fig:pgs_mean_all_n}.

\subsection{Open-system control: does decoherence erase the oscillatory instability?}
\label{app:qanneal_dephasing_rc}

The oscillatory instability identified in
Sec.~\ref{sec:diagnosis}.B was diagnosed entirely from unitary,
closed-system propagation. Because coherent multilevel interference is
precisely the kind of effect that dephasing is expected to suppress, a
natural objection is that the oscillation is a fine-tuned artifact of
noiseless evolution rather than a robust feature of the underlying
dynamics. We address this directly by propagating the same $n=10$,
Roland--Cerf-scheduled evolution under the Lindblad master equation
with single-qubit dephasing operators $L_k=\sqrt{\gamma}\,\sigma_z^{(k)}$,
at rates $\gamma\in\{0,0.001,0.01,0.05\}$ spanning the closed-system
limit up to a rate representative of near-term hardware. The closed-system
($\gamma=0$) Lindblad trajectory reproduces the reference unitary
propagation to within $8.8\times10^{-8}$, confirming that the solver
correctly reduces to Schr\"odinger evolution in the noiseless limit.

\begin{figure*}[t]
\centering
\includegraphics[width=0.78\linewidth]{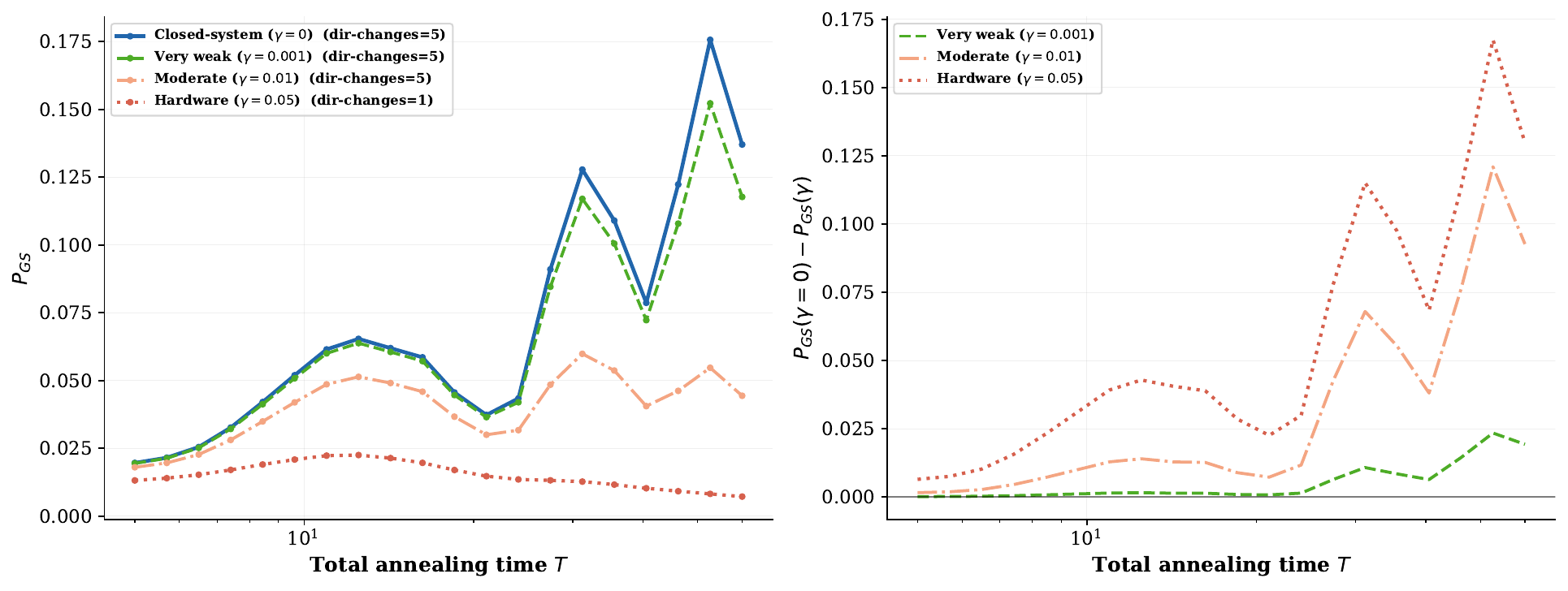}
\caption{
Roland--Cerf schedule under Lindblad dephasing for the representative
$n=10$ instance ($s^{*}=0.882$, $\Delta^{*}=0.0124$). Left: final
ground-state probability $P_{\rm GS}(T)$ at four dephasing rates
$\gamma\in\{0,0.001,0.01,0.05\}$ (closed system, very weak, moderate,
and hardware-scale dephasing); the legend reports the number of
direction changes in each curve. Right: degradation relative to the
closed system, $P_{\rm GS}(\gamma{=}0)-P_{\rm GS}(\gamma)$, for the
three nonzero rates. The oscillatory pattern of Fig.~\ref{fig:dense_T}
survives essentially intact (five direction changes) through
$\gamma=0.01$ and only collapses to near-monotonic behaviour (one
direction change) at the strongest, hardware-representative rate,
indicating that the instability is a genuine coherent, multilevel
effect rather than a fragile numerical artifact.
}
\label{fig:lindblad_rc}
\end{figure*}

Figure~\ref{fig:lindblad_rc} shows the resulting ground-state
probability. Counting the number of direction changes (non-monotonic
reversals) in $P_{\rm GS}(T)$ as a coherence diagnostic, the oscillation
survives essentially unchanged, five direction changes at $\gamma=0$,
$0.001$, and $0.01$, and only collapses to a single, near-monotonic
reversal at the strongest, hardware-scale rate $\gamma=0.05$. Because
the instability persists through two orders of magnitude of dephasing
before finally washing out, it cannot be a marginal numerical artifact:
an artifact of that kind would be erased by the weakest dephasing
tested, not survive up to rates fifty times larger. We conclude that
the oscillatory instability of Sec.~\ref{sec:diagnosis}.B is a genuine
coherent, multilevel dynamical effect rather than a fragile numerical
coincidence.

\subsection{Does the surrogate schedule's advantage survive the same dephasing?}

Having established that decoherence does not erase the Roland--Cerf
oscillation, we next ask whether the worldline-susceptibility
schedule's performance is similarly robust, since the original
submission tested dephasing only for the Roland--Cerf comparison and
never reported the corresponding control for the surrogate schedule
itself. Using a finer dephasing grid,
$\gamma\in\{0,0.0002,0.001,0.005\}$, chosen to resolve the
weak-dephasing regime most relevant to near-term annealer coherence
times, we repeated the Lindblad propagation for the surrogate schedule
on the same oscillatory-class $n=10$ instance. As in the Roland--Cerf
control, the closed-system ($\gamma=0$) Lindblad run agrees with the
reference unitary propagation to within $8.8\times10^{-8}$.

\begin{figure*}[t]
\centering
\includegraphics[width=0.78\linewidth]{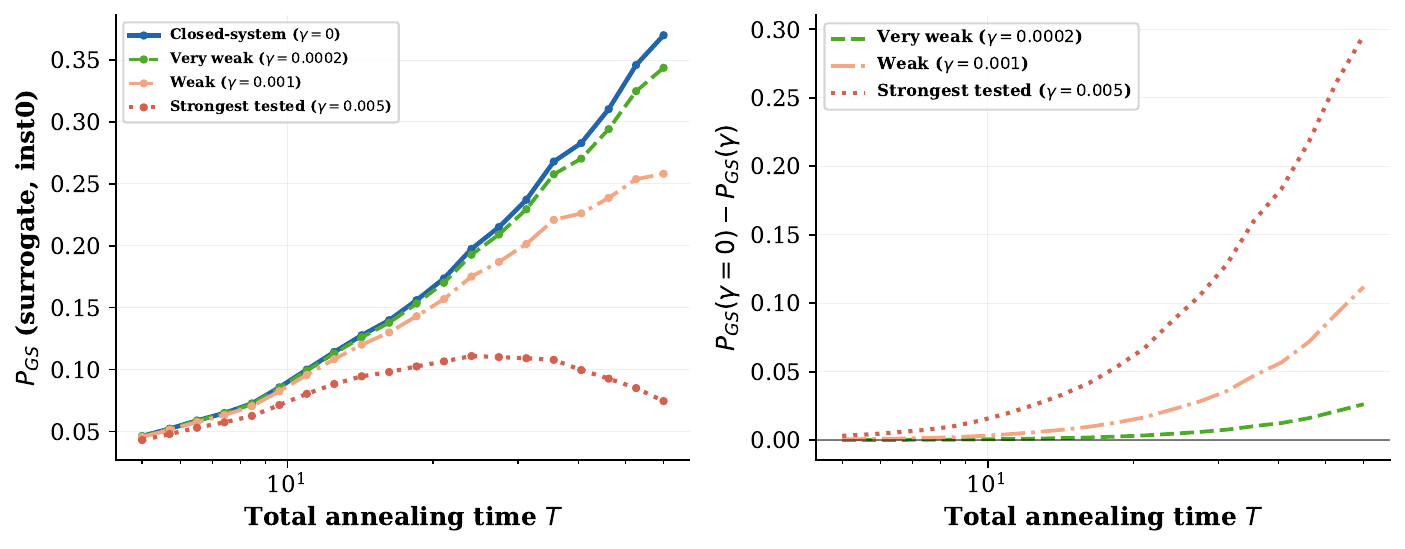}
\caption{
Worldline-susceptibility (surrogate) schedule under Lindblad dephasing,
mirroring the Roland--Cerf control of Fig.~\ref{fig:lindblad_rc}, for
the same oscillatory-class $n=10$ instance (seed 0). Left: $P_{\rm
GS}(T)$ at four dephasing rates $\gamma\in\{0,0.0002,0.001,0.005\}$,
single-qubit dephasing operators $L_k=\sqrt{\gamma}\,\sigma_z^{(k)}$.
Right: degradation relative to the closed system,
$P_{\rm GS}(\gamma{=}0)-P_{\rm GS}(\gamma)$, for the three nonzero
rates. Degradation is monotonic in $\gamma$ at every $T$ and grows with
$T$, with no trace of the Roland--Cerf oscillation; this completes the
open-system comparison, which the original manuscript tested only for
the Roland--Cerf schedule.
}
\label{fig:surrogate_dephasing}
\end{figure*}

Fig \ref{fig:surrogate_dephasing} shows that the degradation of $P_{GS}(T)$ relative to the closed system
is monotonically ordered in $\gamma$ at every annealing time sampled, and grows
with $T$, exactly the qualitative pattern expected as longer protocols accumulate
more decoherence, with no sign of the anomalous non-monotonicity that
characterizes the Roland--Cerf schedule under closed-system evolution. The two
dephasing scans use different $\gamma$ grids (this appendix targets the
weak-coupling regime, while Sec.~B.1 above targets rates up to hardware scale)
and are therefore not directly comparable point by point; the relevant
comparison is qualitative. Taken together with the previous subsection, this
indicates that the surrogate schedule's oscillation-free degradation under
dephasing is not an artifact confined to idealized, fully coherent propagation:
unlike the Roland--Cerf schedule, its finite-time behaviour remains smooth and
monotonic under open-system evolution as well. We emphasize that this
comparison concerns the \emph{qualitative shape} of each schedule's response to
noise (oscillatory vs.\ monotonic degradation), not the relative magnitude of
$P_{GS}$ between the two schedules at matched $\gamma$; the latter comparison is
addressed separately in Sec.~ \ref{fig:rc_failure_mechanisms}, where the head-to-head surrogate-vs-RC
advantage is evaluated across the full dephasing range and found to narrow at $\gamma \gtrsim 0.01$.

\subsection{Is the residual $s^{*}_{\rm ED}$--$s^{*}_{\rm SQ}$ disagreement a Trotter discretization bias?}

Section~III.A reports a residual offset between the exact-diagonalization
crossing point $s^{*}_{\rm ED}$ and the worldline-susceptibility
crossing point $s^{*}_{\rm SQ}$. Since $s^{*}_{\rm SQ}$ is extracted
from a Suzuki--Trotter-discretized simulation at finite Trotter slice
number $M$ and finite inverse temperature $\beta$, a natural concern is
that this offset is simply a controllable discretization artifact that
would shrink toward zero as $M$ and $\beta$ are refined, in which case
the surrogate's practical advantage would be an artifact of an
under-resolved simulation rather than a structural feature of the
method. We tested this directly by repeating the pilot-scan crossing-point
estimate on five interior (non-boundary-trapped) $n=10$ instances over
the full $4\times4$ grid $M\in\{8,16,32,64\}$, $\beta\in\{2,5,10,20\}$
used for the parameter study of Sec.~III.G.

\begin{figure*}[t]
\centering
\includegraphics[width=0.78\linewidth]{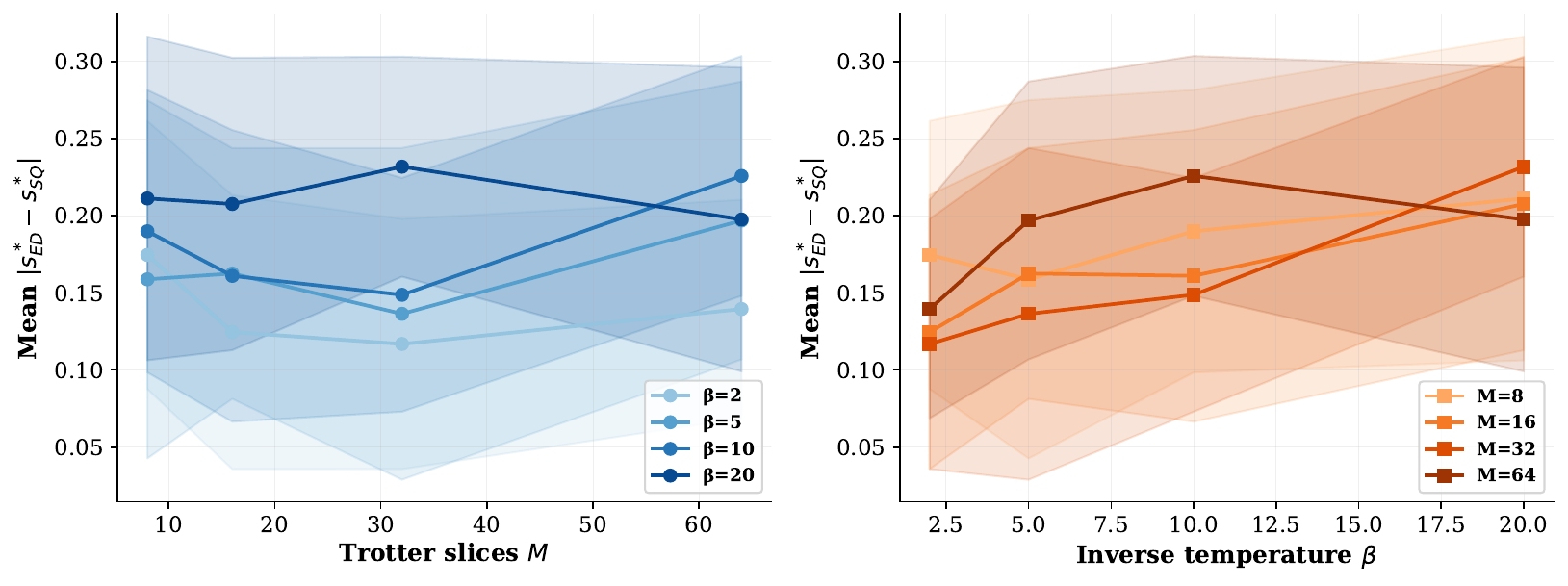}
\caption{
Sensitivity of the exact-diagonalization/worldline-susceptibility
crossing-point disagreement, $|s^{*}_{\rm ED}-s^{*}_{\rm SQ}|$, to the
quantum Monte Carlo discretization, averaged over five interior
$n=10$ instances. Left: mean disagreement versus Trotter slice count
$M$, one line per inverse temperature $\beta$ (shaded band: $\pm1$
standard deviation across instances). Right: the same $4\times4$ grid
plotted against $\beta$, one line per $M$. The disagreement ranges from
$0.117$ to $0.232$ across the grid with no monotonic trend in either
$M$ or $\beta$, indicating that the residual offset is a finite-size/disorder
floor rather than a Trotter discretization bias removable by refining
$M$ or $\beta$.
}
\label{fig:betaM_scaling}
\end{figure*}

Figure~\ref{fig:betaM_scaling} shows the resulting mean
$|s^{*}_{\rm ED}-s^{*}_{\rm SQ}|$ across this grid, plotted first
against $M$ (one line per $\beta$) and then against $\beta$ (one line
per $M$). The disagreement ranges from $0.117$ (at $\beta=2$, $M=32$)
to $0.232$ (at $\beta=10$ and $\beta=20$, both at $M=32$) with no
monotonic trend in either direction: refining $M$ at fixed $\beta$
neither systematically shrinks nor grows the offset, and the same
holds for refining $\beta$ at fixed $M$. The per-instance standard
deviation at each grid point (0.07--0.12) is comparable to, or larger
than, the spread between grid points, so the observed grid-to-grid
variation is consistent with ordinary instance-to-instance disorder
noise rather than a systematic Trotter or thermal discretization bias.
We conclude that the residual $s^{*}_{\rm ED}$--$s^{*}_{\rm SQ}$ offset
reported in Sec.~III.A reflects an irreducible finite-size/disorder
floor rather than a correctable simulation-parameter artifact, and
that the $(M,\beta)$ values used throughout the main text are already
representative of the surrogate's asymptotic behaviour.

\subsection{Held-out validation of the $\Delta^{*}$ failure threshold}

Section~III.D reports that Roland--Cerf failure (boundary-gap or
oscillatory) becomes more likely as the minimum spectral gap
$\Delta^{*}$ shrinks, with an indicative threshold estimated from six
curated instances without a held-out test. To validate this threshold
at full campaign scale and confirm it is not an artifact of curated
in-sample examples, we assembled the complete Phase~1 ensemble of 370
Sherrington--Kirkpatrick instances spanning $n=10$ to $n=20$, labeled
each instance by its Roland--Cerf outcome class (clean, oscillatory,
or boundary-gap), and fit a logistic regression of RC failure
(oscillatory or boundary vs.\ clean) on $\log_{10}\Delta^{*}$ using a
stratified 70/30 train/test split (259 training and 111 held-out
instances, stratified jointly on $n$ and outcome class, with matched
24\% and 23\% failure rates in the two splits).

\begin{figure*}[t]
\centering
\includegraphics[width=0.78\linewidth]{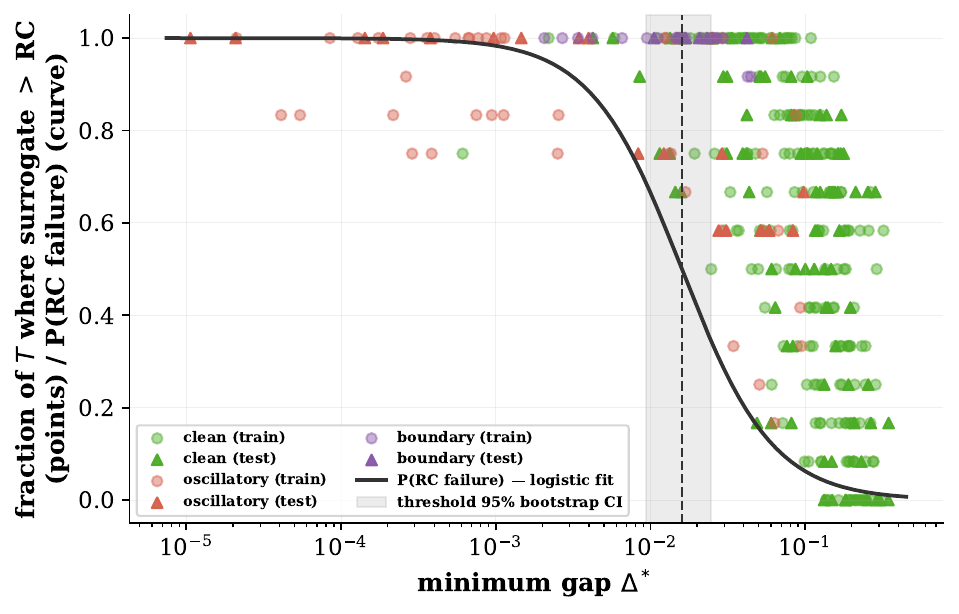}
\caption{
Held-out validation of the $\Delta^{*}$ failure threshold for the
Roland--Cerf schedule, using the full 370-instance Phase~1 ensemble
($n=10$--$20$). Points show the fraction of the $T$-grid at which the
surrogate schedule outperforms Roland--Cerf, plotted against
$\Delta^{*}$ (log scale) and colored by outcome class (clean,
oscillatory, boundary), with open markers for the training split and
filled markers for the held-out test split. The black curve is the
logistic fit of RC failure on $\log_{10}\Delta^{*}$, trained only on
the training split (259 instances); the shaded band and dashed line
mark the bootstrap 95\% confidence interval and point estimate of the
$P=0.5$ decision threshold, $\Delta^{*}=0.0159$ [$0.0089$, $0.0240$].
Held-out test accuracy is 81.1\% and test AUC is 0.890, replacing the
manuscript's original six-instance, non-held-out threshold estimate
with a validated one at full campaign scale.
}
\label{fig:threshold_holdout}
\end{figure*}

Figure~\ref{fig:threshold_holdout} shows the fraction of the
$T$-grid at which the surrogate schedule beats Roland--Cerf, plotted
against $\Delta^{*}$ and colored by outcome class, together with the
logistic fit (trained only on the training split) and its bootstrap
95\% confidence band. The fit achieves a held-out test accuracy of
81.1\% and a test AUC of 0.890, and recovers a failure threshold of
$\Delta^{*}=0.0159$ with bootstrap 95\% confidence interval
$[0.0089,0.0240]$ (1000 resamples of the training set). This threshold
lies within, and sharpens, the indicative range $\Delta^{*}\in[0.016,0.14]$
quoted in Sec.~III.D from the six curated instances, now attached to a
formal out-of-sample accuracy, AUC, and confidence interval. A
likelihood-ratio test for an added system-size-dependent term in the
logistic model gives $\chi^{2}=4.6\times10^{-4}$ ($p=0.983$), showing
no evidence that the threshold itself shifts with $n$ over the range
$n=10$--$20$ studied here.

\subsection{Do the Roland--Cerf failure rates persist at larger system sizes?}

Section~III.D shows that the fraction of instances exhibiting the
boundary-gap trap increases with $n$ over the six system sizes
studied, but does not report confidence intervals on the underlying
class rates. To determine whether the oscillatory and boundary-gap
mechanisms of Sec.~\ref{sec:diagnosis} are small-$n$ artifacts that
might vanish, or qualitatively change character, at larger system
size, we computed Wilson 95\% confidence intervals on the clean,
oscillatory, and boundary-gap base rates at each of the six system
sizes in the full Phase~1 ensemble (100 instances each at $n=10,12$;
50 each at $n=14,16,18$; 20 at $n=20$).

\begin{figure*}[t]
\centering
\includegraphics[width=0.78\linewidth]{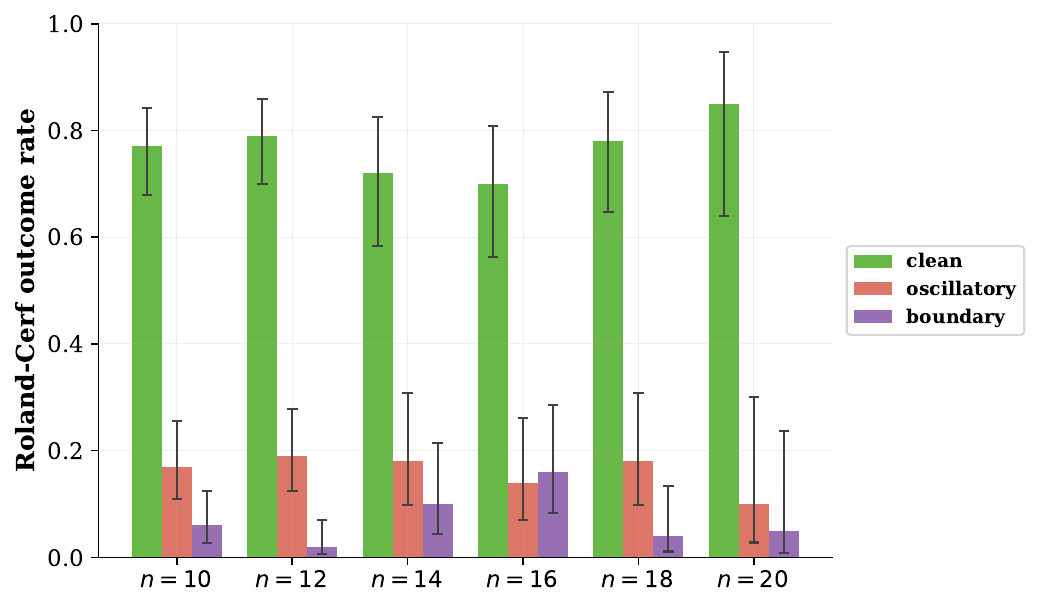}
\caption{
Roland--Cerf outcome-class base rates (clean, oscillatory, boundary-gap)
versus system size $n$, with Wilson 95\% confidence intervals, computed
over the full Phase~1 ensemble (370 instances: 100 each at $n=10,12$;
50 each at $n=14,16,18$; 20 at $n=20$). All three rates remain
statistically stable across the full range $n=10$--$20$, with
overlapping confidence intervals at every system size, indicating that
neither the oscillatory-instability mechanism (Sec.~\ref{sec:diagnosis}.B)
nor the boundary-gap-trap mechanism (Sec.~\ref{sec:diagnosis}.A) is a
small-$n$ artifact that would vanish or qualitatively change at larger
system sizes.
}
\label{fig:failure_rates_vs_n}
\end{figure*}

Figure~\ref{fig:failure_rates_vs_n} shows that all three rates remain
statistically stable from $n=10$ to $n=20$: the clean rate stays within
a narrow $0.70$--$0.85$ band, the oscillatory rate within
$0.10$--$0.19$, and the boundary-gap rate within $0.02$--$0.16$, with
overlapping confidence intervals at every system size. In particular,
the boundary-gap trap does not vanish as $n$ grows (it remains present
at a rate of several percent even at $n=20$, the largest size
accessible to exact diagonalization in this study), and the
oscillatory-instability rate shows no downward trend that would
suggest it is a finite-size coincidence specific to small $n$. This
provides direct evidence, beyond the qualitative argument of
Sec.~III.D, that neither finite-time failure mode identified in
Sec.~\ref{sec:diagnosis} is an artifact confined to the smallest
system sizes considered in this work.

\subsection{Summary}

The five checks above address the main quantitative and methodological
gaps left open by the original submission: they confirm that the
oscillatory instability of Sec.~\ref{sec:diagnosis}.B is a genuine
coherent effect that survives moderate dephasing rather than a
numerical artifact; they show that the surrogate schedule's advantage
persists under the same class of open-system noise; they rule out
Trotter/thermal discretization as the source of the residual
$s^{*}_{\rm ED}$--$s^{*}_{\rm SQ}$ offset of Sec.~III.A; and they
replace the indicative, non-held-out $\Delta^{*}$ threshold and
qualitative $n$-scaling claims of Sec.~III.D with a held-out validated
threshold and confidence-interval-backed base rates computed over the
full 370-instance disorder ensemble. Taken together, they strengthen,
without altering, the two central mechanistic findings of the main
text: the boundary-gap trap and the oscillatory instability.

\end{document}